\documentclass[reprint,twocolumn,superscriptaddress,preprintnumbers,amsmath,amssymb,aps,prd,floatfix]{revtex4-2}

\usepackage[utf8]{inputenc}

\usepackage{mathtools}
\usepackage{amsfonts}
\usepackage{mathrsfs}
\usepackage{bm}
\usepackage[normalem]{ulem}
\usepackage{bbold}
\usepackage{braket}
\usepackage{slashed}
\usepackage{dsfont}
\usepackage{float}
\usepackage{dcolumn}% Align table columns on decimal point
\usepackage{amssymb,amsmath}
\usepackage{titlesec}

\usepackage{graphicx}
\usepackage{color}
\usepackage{array}

\usepackage{placeins}
\usepackage{booktabs}
\usepackage{xspace}
\usepackage{hyperref}
\usepackage[nameinlink]{cleveref}
\usepackage{bookmark}
\usepackage{units}

\setkeys{Gin}{width=0.48\textwidth}

\newcommand{\tr}{{\text{tr}}}

\usepackage{booktabs}
\usepackage{multirow}

\newcolumntype{C}{>{$}c<{$}}
\AtBeginDocument{
	\heavyrulewidth=.08em
	\lightrulewidth=.05em
	\cmidrulewidth=.03em
	\belowrulesep=.65ex
	\belowbottomsep=0pt
	\aboverulesep=.4ex
	\abovetopsep=0pt
	\cmidrulesep=\doublerulesep
	\cmidrulekern=.5em
	\defaultaddspace=.5em
}

\newcommand{\imag}{\text{i}}

\graphicspath{{./figures/}}
\usepackage{xifthen}
\usepackage{xcolor}

\newcommand{\gettitle}{Ripples of the QCD Critical Point}

\hypersetup{
	colorlinks,
	linkcolor={blue!75!black},
	citecolor={blue!75!black},
	urlcolor={blue!75!black}, 
	pdftitle={\gettitle},
	pdfauthor={},
	pdfkeywords={}
	{} 
	bookmarksopen=true,
	bookmarksopenlevel=2,
	bookmarksnumbered=true
}

\begin{document}
\title{\gettitle}

\author{Wei-jie Fu}
\affiliation{School of Physics, Dalian University of Technology, Dalian, 116024, P.R. China}

\author{Xiaofeng Luo}
\affiliation{Key Laboratory of Quark \& Lepton Physics (MOE) and Institute of Particle Physics, Central China Normal University, Wuhan 430079, China}
  
\author{Jan M. Pawlowski}
\affiliation{Institut f\"ur Theoretische Physik, Universit\"at Heidelberg, Philosophenweg 16, 69120 Heidelberg, Germany}
\affiliation{ExtreMe Matter Institute EMMI, GSI, Planckstra{\ss}e 1, D-64291 Darmstadt, Germany}

\author{Fabian Rennecke}
\affiliation{Institut f\"ur Theoretische Physik, Justus-Liebig-Universit\"at Gie\ss en, 35392 Gie\ss en, Germany}
\affiliation{Helmholtz Research Academy Hesse for FAIR, Campus Gie\ss en, 35392 Gie\ss en, Germany}

\author{Shi Yin}
\email{yinshi2014@mail.dlut.edu.cn}
\affiliation{School of Physics, Dalian University of Technology, Dalian, 116024, P.R. China}

\begin{abstract}

We investigate the impact of a critical end point (CEP) on the experimentally accessible baryon number fluctuations of different orders. By now, its potential location has been constrained fairly accurately within first principles functional QCD, together with the location of the chiral crossover line and further thermodynamic observables. This information is incorporated in an advanced QCD-assisted low energy effective theory which is used for the computation of baryon number fluctuations at the chemical freeze-out. This computation also takes care of global baryon number conservation at larger density, where the system changes from grand-canonical to canonical statistics. We observe a prominent peak structure, whose amplitude depends on the location of the CEP, while its position is more sensitive to the location of the freeze-out curve. Our results provide guidance for future low energy heavy-ion experiments.

\end{abstract}

\maketitle

%%%%%%%%%%%%%%%%%%%%%%%%%%%%%%%%%%%%%%%%%%%%%%%%%%%%
\emph{Introduction.--} 
At low densities QCD undergoes a thermal crossover from a deconfined, chirally symmetric high temperature phase to the hadronic phase with confinement and dynamical chiral symmetry breaking at low temperatures. This transition is by now well-understood both experimentally and theoretically. In turn, at high densities this simple crossover behaviour either ends in a critical end point (CEP) or in a different phase, e.g., with spatial modulations \cite{Fukushima:2010bq, Fu:2019hdw, Pisarski:2021qof, Rennecke:2023xhc}. Since it is characterised by the onset of qualitatively different behaviour, we refer to the region where this happens as \emph{onset regime}. Locating this regime and studying its properties via its imprints at chemical freeze out provides the most efficient way to unveil the mysteries of the QCD phase diagram at finite temperature and high baryon densities \cite{Arslandok:2023utm, Luo:2022mtp}.

This onset regime at high densities casts an observable shadow at lower densities: \textit{In the critical regime} close to a CEP, fluctuations of conserved charges show a sizable non-monotonic behaviour due to universal scaling, see e.g.~\cite{Stephanov:1999zu, Stephanov:2008qz, Luo:2017faz, STAR:2020tga, STAR:2021iop, Mroczek:2020rpm, Dore:2022qyz}. It has been shown in \cite{Fu:2021oaw}, that a sizable non-monotonic behaviour is even seen \textit{far outside the critical regime} dominated by  universal scaling. Its origin in this non-critical regime is the successive sharpening of the chiral crossover. Importantly, the details of this sharpening encode the location of the onset regime as ripples in water encode that of the stone causing them. Hence, an in-detail analysis of both experimental measurements and theoretical computations of the observables even far away from this regime can be used to predict or reconstruct its location. This is facilitated due to the emergence of soft (light) modes with a significantly increased correlation length even outside the critical regime \cite{MagEoS:2023}. 

Remarkably, a non-monotonic variation in the kurtosis of net-proton number distributions has recently been observed by the STAR collaboration with $3.1\,\sigma$ significance in the first phase of the Beam Energy Scan (BES) program at the Relativistic Heavy Ion Collider (RHIC) \cite{STAR:2020tga}. The kurtosis of proton distributions at the collision energy $\sqrt{s_{\mathrm{NN}}}=$3\,GeV in the fixed-target experiment is consistent with the hadronic results \cite{STAR:2021fge, STAR:2022etb}, suggesting that a potential CEP is located in a region related to a collision energy above 3\,GeV. 

This is compatible with converged estimates from first-principles functional QCD studies with the functional renormalisation group (fRG) and Dyson-Schwinger equations: the location of the onset regime is constrained by 600\,MeV $\lesssim {\mu_B}_{_{\mathrm{CEP}}}\lesssim$\,650\,MeV, see \cite{Fu:2019hdw, Gao:2020qsj, Gao:2020fbl, Gunkel:2021oya}, and therefore in range of fixed-target experiments. 

A combined in-detail analysis of experimental data and theoretical results opens a path towards pinning down the location and properties of the onset regime. This requires the systematic theoretical study of the physics specific to heavy-ion collisions at lower beam energies as well as their imprints in the experimental data. For that purpose we construct an advanced low-energy effective theory (LEFT) based on state-of-the-art functional QCD results. In this LEFT, the onset is signaled by a CEP. This set-up allows us to unravel experimental imprints of a CEP within baryon number fluctuations at freeze-out. We find a prominent peak structure, whose amplitude depends on the location of the CEP, while its position is more sensitive to the location of the freeze-out curve. In summary, the present approach provides, for the first time, quantitative access to the CEP and its properties. \\[-1.5ex]

%%%%%%%%%%%%%%%%%%%%%%%%%%%%%%%%%%%%%%%%%%%%%
\emph{Functional QCD at finite temperature and density.--}
The phase structure of QCD can be resolved with functional QCD approaches through the non-perturbative computation of QCD correlation functions. These correlation functions allow for the determination of observables such as critical temperatures, condensates and conserved charge distributions.  Importantly, computations at finite density are not obstructed by the sign problem as for lattice QCD, and are only limited by the computational resources required for computations in reliable approximations. By now functional QCD can access the regime with $\mu_B/T \lesssim 4$ with quantitative precision, and allows for qualitative estimates in the regime $\mu_B/T \gtrsim 4$. For recent reviews see \cite{Fischer:2018sdj, Dupuis:2020fhh, Fu:2022gou}. 

These functional QCD studies have passed through strict benchmark tests in the regime of $\mu_B/T\lesssim 2 - 3$, where the respective results on the phase structure and on fluctuations can be compared with results from lattice QCD simulations, see e.g.\ \cite{Bellwied:2015rza, Bazavov:2017dus, Bazavov:2017tot, Bazavov:2018mes, Borsanyi:2018grb, Bazavov:2020bjn, Bollweg:2022fqq} and the recent review \cite{Aarts:2023vsf}. Hence, they represent themselves a self-consistent analytic continuation from QCD with $\mu_B/T\lesssim 3$. Accordingly, results from functional QCD at $\mu_B/T\gtrsim 3$ are not only fully compatible with constraints from analytic extrapolations based on lattice data, see e.g.~\cite{Mukherjee:2019eou, Bollweg:2022rps, Borsanyi:2022soo, Aarts:2023vsf}, their reliability qualitatively surpasses that of analytic extrapolations as they are based on  solving dynamical equations in QCD. 

The present fRG approach with a QCD-assisted LEFT relies on results and technical advances in the description of first-principles QCD at finite temperature and density with the fRG put forward in \cite{Fu:2019hdw}, based on \cite{Braun:2007bx, Braun:2009gm, Mitter:2014wpa, Braun:2014ata, Rennecke:2015eba, Cyrol:2016tym, Cyrol:2017ewj, Cyrol:2017qkl} in the vacuum and at finite temperature. The construction of a quantitatively reliable LEFT is facilitated by the fact that the glue dynamics decouple very efficiently due to the gluonic mass gap of QCD at energy scales of about 1\,GeV \cite{Fu:2019hdw, Dupuis:2020fhh, Fu:2022gou}. This entails that low energy QCD is well described by the respective emergent LEFT: QCD without gluonic fluctuations but in a gluonic background. Moreover, quantitative precision is then obtained by using the fRG to match the RG flows of QCD to those of the LEFT.  This set-up has been named \textit{QCD-assisted LEFT}, and a first study including a detailed discussion of the setup has been presented in \cite{Fu:2021oaw}.

In the present work we aim for quantitative precision and reliable predictions at high densities. This is achieved by directly evolving the RG flow of quark-meson scattering processes obtained in first-principles QCD in \cite{Fu:2019hdw} in our LEFT. These processes encode the correlations between quarks and gluons in the channel that carries the dynamics of the chiral condensate. This allows us to accurately capture the CEP as it arises in QCD at large $\mu_B$, while being in excellent agreement with lattice data at small $\mu_B$. The CEP in the present QCD-assisted LEFT is located at   
\begin{align}
(T_\mathrm{CEP},\mu_{B_{\mathrm{CEP}}})=(98,643)\,{\rm MeV}\,,
\end{align}
consistent with the constraint  
\begin{align}
600\,\textrm{MeV} \lesssim {\mu_B}_{_{\mathrm{CEP}}}\lesssim 650\, \textrm{MeV}\,,
\label{eq:CEPEstimateQCD}
\end{align}
in full functional QCD, \cite{Fu:2019hdw, Gao:2020qsj, Gao:2020fbl, Gunkel:2021oya}. \Cref{eq:CEPEstimateQCD} is the state-of-the-art uncertainty estimate for the location of CEP from functional QCD. Variations of the CEP location within this regime are possible and we shall use them later for an investigation of the experimental imprints and properties of the regime around the CEP. The details of our setup can be found in the supplement. \\[-1.5ex]

%%%%%%%%%%%%%%%%%%%%%%%%%%%%%%%%%%%%%%%%%%
\emph{Baryon number fluctuations at freeze-out.--} 
In the present work we use vanishing chemical potentials for the electric charge and strangeness, as the effects of the corresponding charge conservation are subleading for baryon-number fluctuations \cite{Fu:2018qsk, Fu:2018swz, Wen:2019ruz}. We thus compute the grand potential $\Omega[T, \mu_B]$ and extract from it the pressure,
\begin{align}
    p=&-\Omega[T,\mu_B]\,,
    \label{eq:pres}
\end{align}
and the generalised susceptibilities,
\begin{align}
    \chi_{n}^{B}=&\frac{\partial^{n}}{\partial(\mu_{B}/T)^{n}}\frac{p}{T^{4}}\,. 
    \label{eq:chi-def}
\end{align}
The $\chi_{n}^{B}$ are directly related to the cumulants of the net-baryon number distribution, whose proxy, the net-proton distribution, can be measured in the experiments \cite{Hatta:2003wn}. The cumulants of the lowest four orders, the mean value $M$, the variance $\sigma$, the skewness $S$ and the kurtosis $\kappa$, are given by  
\begin{align}
 \frac{M}{VT^3}=\chi_1^{B}\,,\ \frac{\sigma^2}{VT^3}=\chi_2^{B}\,,\   
S=\frac{\chi_3^{B}}{\chi_2^{B}\sigma}\,,\  \kappa=\frac{\chi_4^{B}}{\chi_2^{B}\sigma^2}\,, 
\end{align}
where we have already divided out the volume dependence. The latter is naturally absent in the ratio between two susceptibilities of different orders, 
\begin{align}
    R_{nm}^{B}&=\frac{\chi_n^{B}}{\chi_m^{B}}\,.
    \label{eq:Rnm}
\end{align}
These ratios have been computed in equilibrium and at vanishing density in lattice QCD, e.g.~\cite{Bazavov:2012vg, Borsanyi:2013hza, Borsanyi:2014ewa, Bazavov:2020bjn} and with functional methods both at vanishing and finite density, e.g.~\cite{Fu:2015naa, Fu:2015amv, Fu:2016tey, Almasi:2017bhq, Isserstedt:2019pgx, Fu:2021oaw}. 

In particular in the regime of low collision energy, high-order baryon number fluctuations are significantly suppressed by global baryon number conservation, \cite{Bzdak:2012an, Vovchenko:2020tsr, Braun-Munzinger:2020jbk}. In order to accurately describe the relevant features of the medium created in heavy-ion collisions, this is taken into account here by considering canonical corrections to grand canonical susceptibilities. To this end, we adopt the subensemble acceptance method (SAM) as proposed in \cite{Vovchenko:2020tsr}. In SAM the ratio between the subensemble volume, $V_1$, measured in the acceptance window and that of the whole system, $V$, is given by $\alpha=V_1/V$. In the thermodynamic limit, where both sizes of total- and sub-systems are significantly larger than the correlation length $\xi$, the measured cumulants in the sub-system approach the grand canonical values discussed above when $\alpha \to 0$. When the effect of global baryon number conservation begins to play a role, the parameter $\alpha$ develops a nonzero value and canonical corrections apply. 

We fix $\alpha$ with the most sensitive and well-observed ratio of low-order fluctuations. This is $R_{32}$, for which the experimental data show a significant flattening for $\sqrt{s_{\mathrm{NN}}}\lesssim$\,11.5\,GeV which is not seen in $R_{32}$ computed within the grand canonical ensemble. We attribute a sizable part of it to the increasing importance of canonical statistics and use the discrepancy to fix $\alpha$. Notably, there are further effects such as the detector acceptance \cite{Luo:2017faz} and volume fluctuations \cite{Braun-Munzinger:2016yjz}. They are considered to be subleading in the present work. Moreover, they may also be partially accounted for by our choice of $\alpha$. 

\begin{figure}[t]
\includegraphics[width=0.4\textwidth]{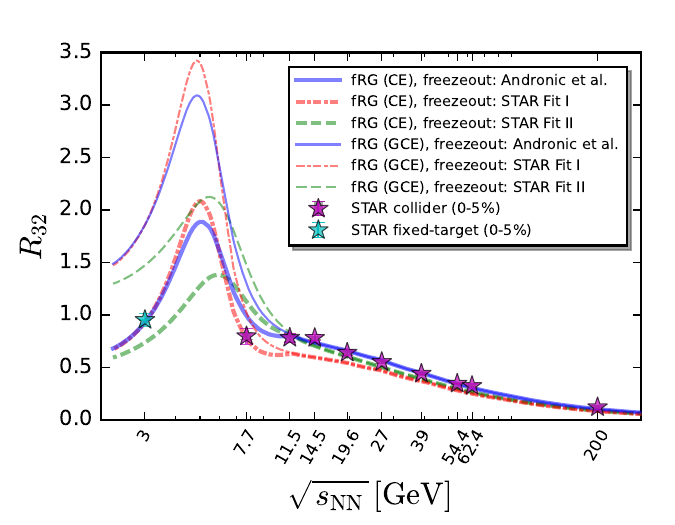}
\caption{Ratio $R_{32}$ of baryon number fluctuations defined in \labelcref{eq:Rnm} for the grand canonical ensemble (GCE) and with canonical corrections at small beam energies due to global baryon number conservation (CE). The latter are taken into account with SAM \cite{Vovchenko:2020tsr} and fixed using STAR data from Refs.\ \cite{STAR:2020tga, STAR:2021fge,STAR:2021iop}. Three different freeze-out curves are used \cite{Fu:2021oaw}.}
\label{fig:R32-alpha}
\end{figure}

In \Cref{fig:R32-alpha} we depict our theoretical results and the experimental measurements for $R_{32}$. Our results are shown both without and with canonical corrections. Evidently, for $\sqrt{s_{\mathrm{NN}}}\gtrsim$\,11.5\,GeV, the effects of global baryon number conservation are negligible and we use $\alpha=0$ in this range. For $\sqrt{s_{\mathrm{NN}}}\lesssim$\,11.5 GeV the importance of global baryon number conservation increases. The qualitative behaviour is well captured with a linear dependence of the parameter $\alpha$ on $\sqrt{s_{\mathrm{NN}}}$, to wit, 
\begin{subequations}
\label{eq:alpha}
\begin{align}
  \alpha(\bar s) = a \left(1- \sqrt{\bar s}\right) \theta\left(1-\bar s \right) \,,
  \label{eq:alphaFun}
\end{align}
with the Heaviside $\theta$-function and the parameters 
\begin{align}
a=0.33\,, \qquad \sqrt{\bar s}= \frac{\sqrt{s_{\mathrm{NN}}}}{11.9\,\textrm{GeV}}\,.
\label{eq:alphaParameters}
\end{align}
\end{subequations}
The parameters are obtained by matching our results with the experimental data at three different energies. We find $\alpha=0$ at $\sqrt{s_{\mathrm{NN}}}=$\,11.5\,GeV, $\alpha=0.14$ at $\sqrt{s_{\mathrm{NN}}}=$ 7.7 GeV, and $\alpha=0.24$ at $\sqrt{s_{\mathrm{NN}}}=$ 3 GeV. Fitting these points yields the parameters in \labelcref{eq:alpha}. More details are provided in the supplement. 

%
%%%%%%%%%%%%%%%%%%%%%%%%%%%%%
%%%%%%%%%%%%%%%%%%%%%%%%%%%%%
%

Finally, we use three different freeze-out curves introduced in our previous work \cite{Fu:2021oaw}: The first one is obtained through a parametrisation of the freeze-out data from \cite{Andronic:2017pug}; the others are based on the data from the STAR experiment \cite{Adamczyk:2017iwn}, and are denoted by STAR Fit I and STAR Fit II. In STAR Fit I, all the freeze-out data from STAR are used in the fit. In STAR Fit II some collision energy data are dropped, considered to be flawed from general considerations. For more details on the fitting procedure see \cite{Fu:2021oaw}. \\[-1.5ex]

\begin{figure}[t]
\includegraphics[width=0.5\textwidth]{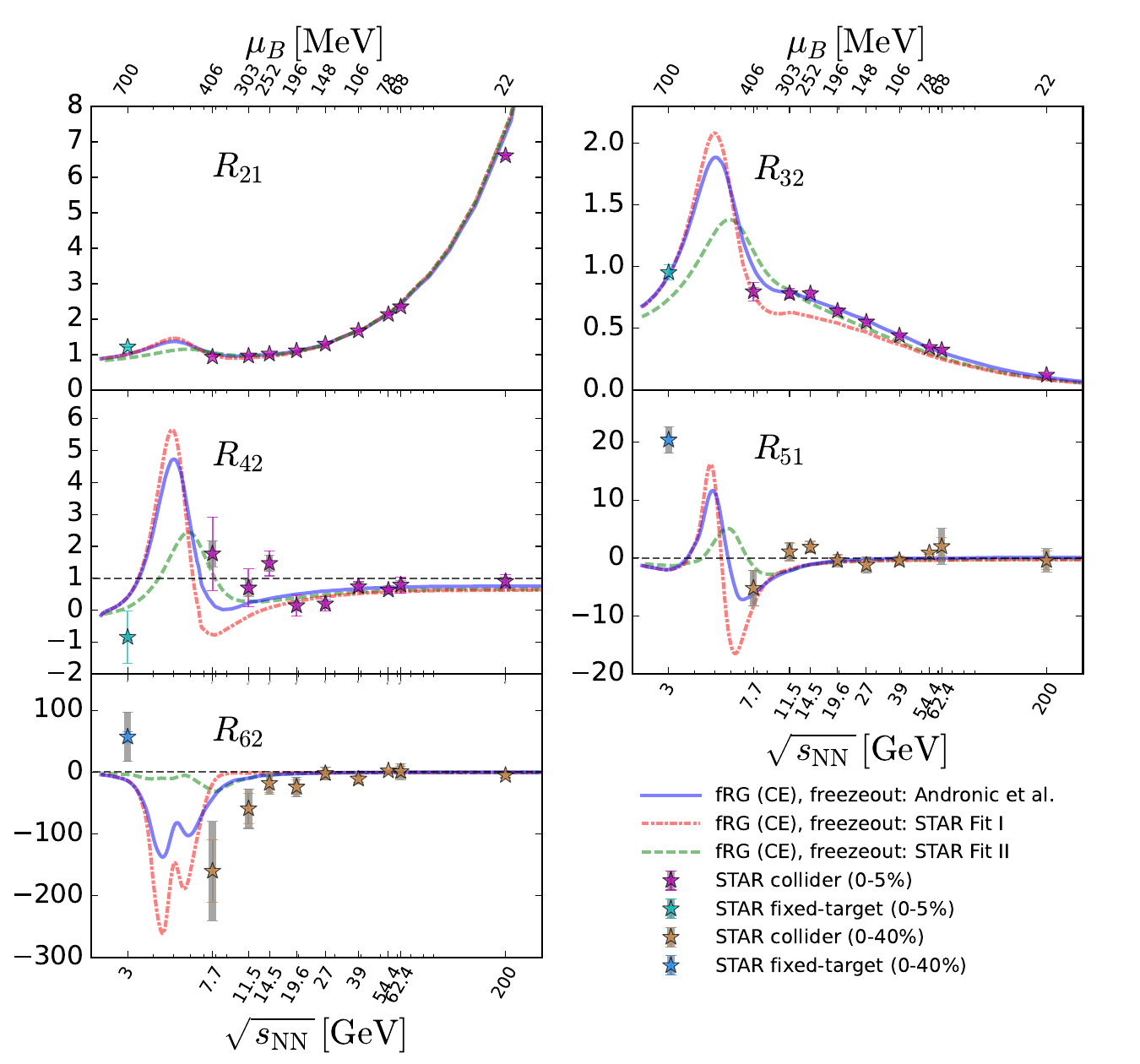}
\caption{Baryon number fluctuations of different orders as functions of the collision energy, calculated in the QCD-assisted LEFT with the fRG on three different freeze-out curves \cite{Fu:2021oaw}, where the effects of global baryon number conservation are taken into account in the range of $\sqrt{s_{\mathrm{NN}}}\lesssim$\,11.5\,GeV through SAM \cite{Vovchenko:2020tsr}, see text for more details. Experimental data measured by STAR are also presented for comparison~\cite{STAR:2020tga, STAR:2023esa, STAR:2021rls,STAR:2021fge,STAR:2021iop}.}
\label{fig:chi-CE}
\end{figure}

%%%%%%%%%%%%%%%%%%%%%%%%%%%%%%%%%%%%%%%%%%%%%%%%%%%
\emph{Results for baryon number fluctuations.--} 
With our setup in place, we can investigate further fluctuation observables. In \Cref{fig:chi-CE} we show the ratios of baryon number fluctuations $R_{21}^{B}$, $R_{32}^{B}$, $R_{42}^{B}$, $R_{51}^{B}$, $R_{62}^{B}$ with canonical corrections as functions of the collision energy for the three different freeze-out curves mentioned above. Our results are compared with experimental measurements by the STAR collaboration, including the ratio between the variance and mean value $R_{21}^p$, skewness $R_{32}^p$, kurtosis $R_{42}^p$ of net-proton distributions for central (0-5\%) Au+Au collisions \cite{STAR:2020tga}, fifth- and sixth-order net-proton fluctuations $R_{51}^p$, $R_{62}^p$ with centrality 0-40\% \cite{STAR:2023esa}, proton cumulants in fixed-target collisions at $\sqrt{s_{\mathrm{NN}}}$=3 GeV \cite{STAR:2021fge}.

%
%%%%%%%%%%%%%%%%%%%%%%%%%%%%%
%%%%%%%%%%%%%%%%%%%%%%%%%%%%%
% 

In general, \Cref{fig:chi-CE} shows that our results are in very good agreement with the experimental data, keeping in mind the caveats of directly comparing conserved charge and net-proton fluctuations \cite{Luo:2017faz, Bzdak:2019pkr, Vovchenko:2021kxx}. For $R_{21}$ we see a small deviation at the highest collision energy $\sqrt{s_{\mathrm{NN}}}$=200 GeV. This can be attributed to a slight mismatch in the freeze-out chemical potentials at this beam energy, namely ${\mu_B}_{_{\mathrm{CF}}}\simeq 22$\,MeV in theory and ${\mu_B}_{_{\mathrm{CF}}}\simeq 27$\,MeV in experiment \cite{Adamczyk:2017iwn}. $R_{42}^{p}$ is perhaps the most prominent experimental observable used in the search for the CEP in the QCD phase diagram. The significant non-monotonic behaviour for 7.7\,GeV $\leq\sqrt{s_{\mathrm{NN}}}\leq$ 200\,GeV observed by STAR \cite{STAR:2020tga} is also seen in our results. For the hyper-order fluctuations $R_{51}^{B}$ and $R_{62}^{B}$ we find that the fRG results are qualitatively consistent with experimental data with $\sqrt{s_{\mathrm{NN}}}\gtrsim$ 7.7 GeV, where both $R_{51}^{B}$ and $R_{62}^{B}$ develop negative values with decreasing collision energy. In comparison to the grand canonical results presented in the supplement, one finds that the magnitudes of high-order fluctuations, especially the hyper-order ones, are suppressed considerably in the regime of low collision energy. Note that both values of $R_{62}^{B}$ and $R_{51}^{B}$ at $\sqrt{s_{\mathrm{NN}}}$=3 GeV are negative, which seems to be inconsistent with experimental data. However, their data is taken at 0-40\% centrality \cite{STAR:2023esa}, while our theoretical description is not geared towards off-central collisions.

%
%%%%%%%%%%%%%%%%%%%%%%%%%%%%%
\begin{figure}[t]
\includegraphics[width=0.4\textwidth]{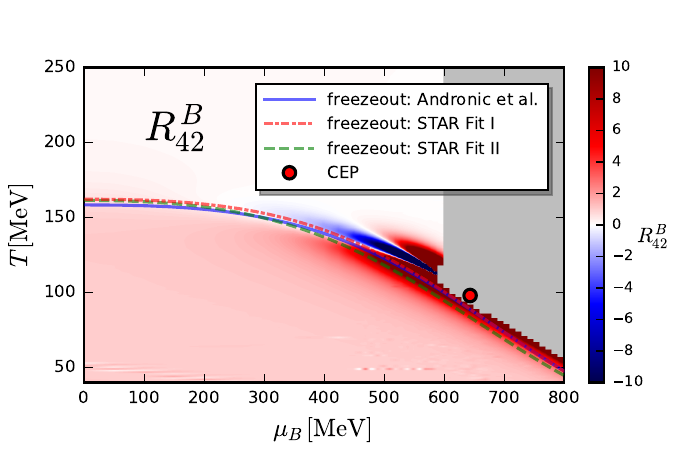}\\ \vspace{-0.5cm}
\includegraphics[width=0.4\textwidth]{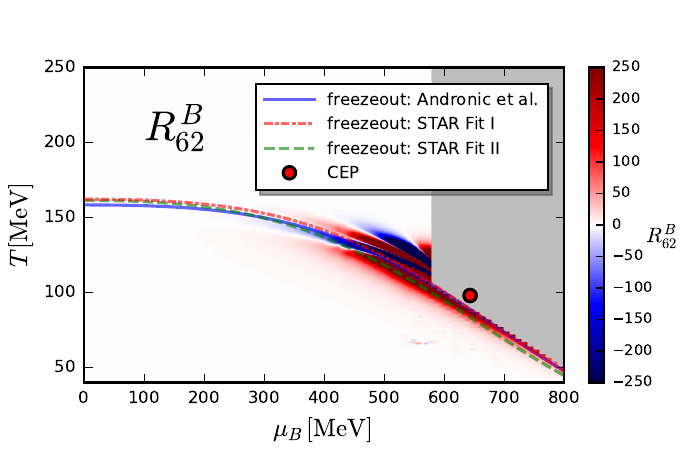}
\caption{Heatmap of $R_{42}^{B}$ (top) and $R_{62}^{B}$ (bottom) in a grand canonical ensemble in the phase diagram obtained in the QCD-assisted LEFT within the fRG with a CEP located at $(T_\mathrm{CEP},\mu_{B_{\mathrm{CEP}}})=(98,643)$\,MeV, where three different freeze-out curves are depicted \cite{Fu:2021oaw}. The gray area depicts the region where the  computation of $R_{42}^{B}$ or $R_{62}^{B}$ is inaccessible within the present numerical set-up.}
\label{fig:R42R62-phasediagram}
\end{figure}
%%%%%%%%%%%%%%%%%%%%%%%%%%%%%
%

A common feature of all observables shown in \Cref{fig:chi-CE} is a pronounced peak structure between 3 and 7.7\,GeV. These peaks appear to be quite sensitive to the choice of freeze-out curve. Additional experimental data in this region would therefore greatly constrain the location of the freeze-out curve at small beam energies. Still, this peak is present for all freeze-out curves and, as shown in the supplement, persists under a variation of the location of the CEP. 

The connection between the cumulants in the phase diagram and the freeze-out can be read-off from \Cref{fig:R42R62-phasediagram}, where we show the phase diagram of the QCD-assisted LEFT together with a heat map of $R_{42}^{B}$ (top panel) and  $R_{62}^{B}$ (bottom panel) in the plane. One can clearly see that the non-monotonic beam-energy dependence of these ratios and the peak structure both arise from the freeze-out line crossing through positive and negative regions for the ratios. These regions are concentrated in a narrow region around the chiral crossover and more pronounced they become, the sharper the crossover is. For example, the non-monotonic behavior and the peak structure of $R_{42}^{B}$ for $\sqrt{s_{\mathrm{NN}}}\lesssim$ 7.7 GeV can be explained by the freeze-out curve moving away from the transition line, thus it deviates from the negative region and crosses over the lower positive region. The freeze-out baryon chemical potentials of the peak are around 536, 541 and 486\,MeV for the freeze-out curves Andronic {\it et al.}, STAR Fit I and STAR Fit II, respectively. Hence, they are at significantly smaller chemical potentials than the location of the CEP at $\mu_{B_{\mathrm{CEP}}}=643$\,MeV. However, the crossover is already quite sharp at these chemical potentials, which leads to large amplitudes of the ratios and hence pronounced peaks.

%
%%%%%%%%%%%%%%%%%%%%%%%%%%%%%
\begin{figure}[t]
\includegraphics[width=0.35\textwidth]{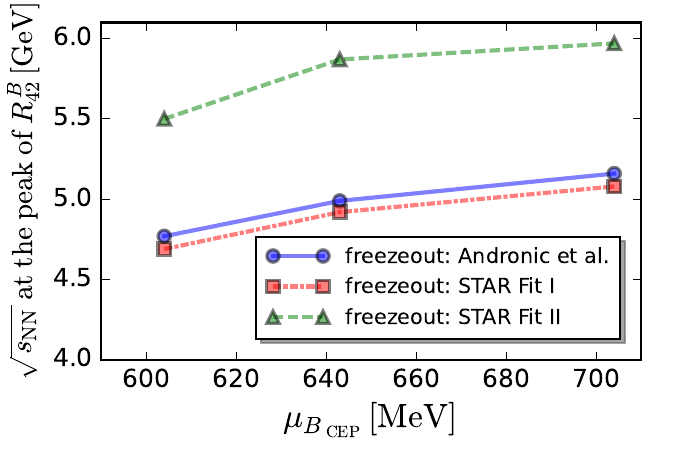}\\ \vspace{-0.14cm}
\includegraphics[width=0.35\textwidth]{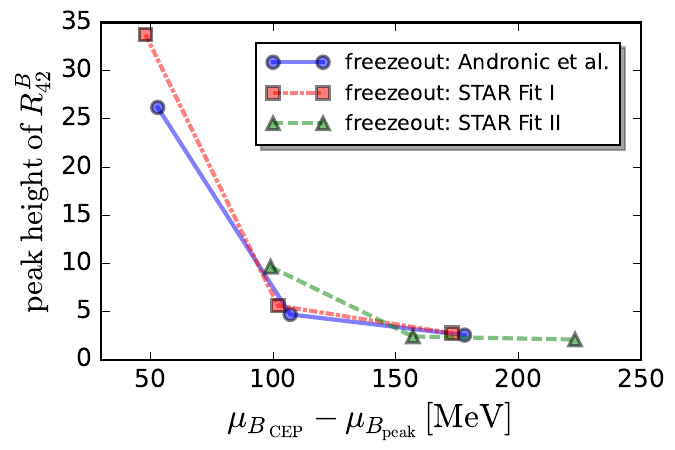}
\caption{Top: Dependence of the position of the peak in $R_{42}^{B}$ on the location of the CEP and the freeze-out curve \cite{Fu:2021oaw}. Bottom: Height of the peak in $R_{42}^{B}$ as a function of the difference between the ${\mu_B}_{_{\mathrm{CEP}}}$ and  ${\mu_B}_{_{\mathrm{peak}}}$, where the latter corresponds to $\mu_B$ related to the $\sqrt{s_{\mathrm{NN}}}$ of the peak in $R_{42}^{B}$.}
\label{fig:peak-muBCEP}
\end{figure}
%%%%%%%%%%%%%%%%%%%%%%%%%%%%%
%

These findings beg the question as to what determines the height and position of the peak. This can be analysed by varying the location of the CEP in the QCD-assisted LEFT within the uncertainty band in full functional QCD in  \labelcref{eq:CEPEstimateQCD}. We use set-ups  with CEPs at $(T_{_{\mathrm{CEP}}},{\mu_B}_{_{\mathrm{CEP}}})=(108, 604)$ \,MeV and at $(94, 704)$\,MeV, and investigate the imprints of this variation on the susceptibilities. The underlying procedure is detailed in the supplement. We find that the position of the peak, i.e., the beam energy $\sqrt{s_{\mathrm{NN}}}$ at the peak in $R_{42}^B$, is rather insensitive to the location of the CEP, and only depends on the freeze-out curve, see the top panel of \Cref{fig:peak-muBCEP}. In order to reduce the uncertainty of the freeze-out curve, in the bottom panel of \Cref{fig:peak-muBCEP} we depict the height of the peak as a function of  ${\mu_B}_{_{\mathrm{CEP}}}-{\mu_B}_{_{\mathrm{peak}}}$, where ${\mu_B}_{_{\mathrm{peak}}}$ is the chemical potential related to $\sqrt{s_{\mathrm{NN}}}$ of the peak. The results of three different freeze-out curves almost fall into one single curve. Most importantly, the closer the CEP is to the peak, the higher it is. For ${\mu_B}_{_{\mathrm{CEP}}}-{\mu_B}_{_{\mathrm{peak}}}\lesssim 100$\,MeV, the peak height increases significantly. In short, the information on the location of the CEP is predominantly encoded in the peak height. \\[-1.5ex]

%%%%%%%%%%%%%%%%%%%%%%%%%%%%%%%%
\emph{Conclusions and summary.--}
In the present work we have presented a satisfactory description of the experimentally measured baryon number fluctuations. In particular, we predict a characteristic peak structure between beam energies of 3 and 7.7\,GeV. The peak can be traced back to a sharpening of the chiral crossover with increasing chemical potential, which is a solid prediction of functional methods \cite{Fu:2019hdw, Gao:2020qsj, Gao:2020fbl, Gunkel:2021oya}. 

Our main finding is that the height of this peak is sensitive to the location of the CEP, while its location depends on the details of the freeze-out. The latter occurs in a regime where the $R_{nm}^B$ considered here, show no sign of critical scaling. Hence the freeze-out happens outside the universal scaling regime of the CEP in terms of these observables. This has the important implication that our predictions are not affected by non-equilibrium effects that arise from critical slowing down \cite{Berdnikov:1999ph, Mukherjee:2016kyu}.
However, since the height of the peak in the kurtosis carries sensitive information on the location of the CEP, the ripples of the CEP can be used to infer its location without the need to rely on critical scaling.

Note that the present set-up can also accommodate more general scenarios for the onset regime in future investigations. In  these scenarios the crossover may end in a spatially modulated regime and not a CEP, e.g.~a moat regime \cite{Fu:2019hdw,Pisarski:2021qof,Rennecke:2023xhc} or even an inhomogeneous phase \cite{Fukushima:2010bq, Buballa:2014tba}, whose detection asks for even more experimental precision data.

A more quantitative in-detail analysis with the present combined experimental-theoretical approach will allow us to answer both, the question of the precise location of the onset regime and its properties. The success of the endeavor of unravelling the QCD phase structure at high baryon densities requires experimental precision data in this regime, which highlights the importance and discovery potential of future heavy-ion  experiments. \\[-1.5ex] 

%%%%%%%%%%%%%%%%%%%%%%%%%%%%
\emph{Acknowledgements.--}
We thank the members of the fQCD collaboration \cite{fQCD} for discussions and collaborations on related projects. This work is funded by the National Natural Science Foundation of China under Grant Nos. 12175030, 12122505, 11890711, and the National Key Research and Development Program of China under contract Nos. 2022YFA1604900, 2020YFE0202002, 2018YFE0205201, and the Deutsche Forschungsgemeinschaft (DFG, German Research Foundation) under Germany’s Excellence Strategy EXC 2181/1 - 390900948 (the Heidelberg STRUCTURES Excellence Cluster) and the Collaborative Research Centre SFB 1225 - 273811115 (ISOQUANT), and the Collaborative Research Centre  TransRegio CRC-TR 211 "Strong-interaction matter under extreme conditions"-- project number 315477589 -- TRR 211.

\bibliography{ref-lib}% Produces the bibliography via BibTeX.

%%%%%%%%%%%%%%%%%%%%%%%%%%%%%%%%%%%%%%%%%%%%%
\newpage

%\appendix 
\renewcommand{\thesubsection}{{S.\arabic{subsection}}}
\setcounter{section}{0}
\titleformat*{\section}{\centering \Large \bfseries}

\onecolumngrid

%\begin{widetext}
%	\newcounter{totalequations}
\section*{Supplemental Materials}
%	\parttoc % Insert the appendix TOC
The supplemental materials provide some details of our theoretical set-up, \Cref{app:setup}, estimates of location of the critical end point (CEP) in the QCD phase diagram, \Cref{app:CEP-Estimates}, a comparison of the results within the canonical ensemble to that within the grand canonical ensemble, \Cref{app:GCE}, details on the subensemble acceptance method (SAM), \Cref{app:canonical}, and finally a detailed study of the imprint of the location of the CEP in baryon number fluctuations, \Cref{app:locCEP}.

%%%%%%%%%%%%%%%%%%%%%%%%%%%%%%%%%%%%%%%%%%
\subsection{Setup of the QCD-assisted LEFT}
\label{app:setup}

In this work we adopt a low-energy effective field theory for the quarks and mesons embedded in the functional QCD set-up in \cite{Fu:2019hdw}. The present version is a systematic improvement towards full QCD of that used in our previous work \cite{Fu:2021oaw}. The respective scale-dependent effective action  reads 
\begin{align}
    \Gamma_k=&\int_x \Biggl\{Z_{q,k}\,\bar{q} \Big [\gamma_\mu \partial_\mu -\gamma_0(\hat\mu+i g A_0) \Big ]q+\frac{1}{2}Z_{\phi,k}(\partial_\mu \phi)^2 +h_k\,\bar{q}\left(T^0\sigma+i\gamma_5\bm T\cdot\bm{\pi}\right)q +V_k(\rho,A_0)- c_\sigma \sigma \Biggr\}\,,
    \label{eq:action}
\end{align}
with the infrared cutoff or renormalisation group (RG) scale $k$, $\int_{x}=\int_0^{1/T}d x_0 \int d^3 x$ and $(T^0,\bm T)=1/2 (\mathbb{1}, \bm \sigma)$, where $\bm \sigma$ denotes the Pauli matrices. In \labelcref{eq:action} the quark field of two light flavors $q=(u,\,d)$ interacts with the mesonic field $\phi=(\sigma, \bm\pi)$ via a $k$-dependent Yukawa coupling $h_k$, where $\sigma$ and $\bm\pi$ are the scalar sigma and pseudoscalar pion fields, respectively. Here, $Z_{q,k}$ stands for the wave function of the quark field and $Z_{\phi,k}$ that of the meson field. The quark chemical potential matrix in the flavor space reads $\hat\mu = \mathrm{diag}(\mu,\mu)$ with $\mu=\mu_B/3$. The masses and interactions of mesons are encoded in the matter part of the effective potential $V_k(\rho,A_0)$, cf. \labelcref{eq:Vtotal}. The potential is O(4) invariant with $\rho = \phi^2/2$. The linear term in the sigma field breaks the chiral symmetry explicitly, and the strength is controlled by the coefficient $c_\sigma$. In the regime of low energy the glue sector, including the gluons and ghosts in the Landau gauge, decouples from the matter sector, since a finite mass gap of gluons develops in this region, for recent reviews see e.g. \cite{Papavassiliou:2022wrb, Fischer:2018sdj, Dupuis:2020fhh, Fu:2022gou}. In the functional renormalisation group (fRG) approach to QCD this has been implemented in \cite{Mitter:2014wpa, Braun:2014ata, Cyrol:2016tym, Cyrol:2017ewj, Fu:2019hdw}.

%%%%%%%%%%%%%%%%%%%%%%%%%%%%%%%%%%%%%%%%%%
\subsubsection{Polyakov loop potential}
\label{app:PolPot}

At low momentum scales the glue dynamics decouples due to the gluonic mass gap, and diagrams with gluon lines get increasingly irrelevant. Still, the on-shell gluonic background contributes, which is reflected by the dropping expectation value of the Polyakov loop $L[A_0]$,  
\begin{align} 
    L(A_0)=\frac{1}{N_c} \Big\langle \tr_f\, P(\bm x)\Big\rangle\,,  \qquad \bar L(A_0)=\frac{1}{N_c} \Big\langle \tr_f\, P^\dagger(\bm x)\Big\rangle\,, \quad \textrm{with} \quad P(\bm x) ={\cal P} \exp \left(\imag \int_0^\beta d\tau A_0(\tau,\bm x)\right)\,, \label{eq:DefofLbarL}
\end{align}
where ${\cal P}$ stands for the path ordering of the exponential and $\tr_f$ is the color trace in the fundamental representation. In the present functional approach to QCD the Polyakov loop is captured by a constant temporal gluon background field $A_0$ (or rather its gauge invariant eigenvalues) at finite temperature and density, 
see \cite{Braun:2007bx, Fister:2013bh}. 

\begin{figure*}[t]
\includegraphics[width=0.5\textwidth]{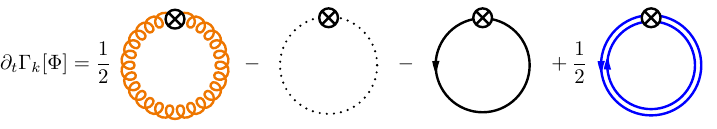}
\caption{Flow equation for the QCD effective action $\Gamma[\Phi]$, where $\Phi$ comprises all fields, see \labelcref{eq:Phi}. Here $t=\log k/\Lambda$ is the (negative) RG-time with some reference scale $\Lambda$ that is absent in the correlation functions. The spiraling orange line denotes a full (momentum- and mean field-dependent) gluon propagator, the black dotted line a full ghost propagator, the straight black line a full quark propagator and the straight blue double line depicts pion and $\sigma$-mode propagators, that are included in the fRG approach to QCD as emergent composites (dynamical hadronisation).}
\label{fig:QCD_FlowEquation}
\end{figure*}

The temperature and density dependent value of the background $A_0$ and hence the Polyakov loop is determined from the effective potential $V(\rho,A_0)$, that is 
given by the QCD effective action $\Gamma[\Phi]$, evaluated on constant backgrounds. Here, $\Phi$ comprises all fields in the effective action, 
\begin{align}
    \Phi=(A,c,\bar c,q, \bar q, \sigma,\bm{\pi})\,.
    \label{eq:Phi}
\end{align}
The potential $V(\rho,A_0)$ can be directly determined from the flow of the QCD effective action depicted in \Cref{fig:QCD_FlowEquation}, evaluated on constant background. We note that the flow of the current low energy effective theory is embedded in the full QCD flow which is described in \labelcref{app:Yukawa}. 
\Cref{fig:QCD_FlowEquation} also entails that the effective potential is a sum of the pure glue part and the matter part, i.e.,
\begin{align}
V_k(\rho,A_0)=& V_{\mathrm{glue},k}(A_0)+V_{\mathrm{mat},k}(\rho,A_0)\,. 
\label{eq:Vtotal}
\end{align}
The glue part of the potential is the $t$-integrated sum of the gluon and ghost diagrams in \Cref{fig:QCD_FlowEquation}, while the matter part is the integrated sum of the quark and meson loops. Here $t$ is the (negative) RG-time $t=\log k/\Lambda$ with some reference scale $\Lambda$. 
%
%%%%%%%%%%%%%%%%%%%%%%%%%%%%%
%%%%%%%%%%%%%%%%%%%%%%%%%%%%%
%

The potential \labelcref{eq:Vtotal} only depends on the Cartan components $A_0^{3,8}$ of the gauge field, the coefficients of the generators $\lambda^{3,8}/2$ with the Gell-Mann matrices $\lambda^a$. This can be shown by using the gauge invariance of the $V(A_0,\rho)$ for rotating the constant background $A_0$ into the Cartan. In general, $L\,,\,\bar L$ are nontrivial functions of the expectation values $A_0^{3,8}$, i.e., $L=L(A_0^3, A_0^8)$ and $\bar L=\bar L(A_0^3, A_0^8)$. Finally, we note that in general $L(A_0)\neq  \tr_f P(A_0)/N_c$, and mapping the $A_0$-solution to the equations of motion, $\langle A_0\rangle$, to $L,\bar L$ within the fRG has been discussed and benchmarked in \cite{Herbst:2015ona} in pure Yang-Mills theory: The respective $L,\bar L$ computed from $A_0$ are in quantitative agreement with lattice results. 

This sets the stage for the computation of the effective $A_0$-potential in the present work. Specifically, a formulation in terms of $L\,,\,\bar L$ provides a direct relation to lattice results, as \labelcref{eq:DefofLbarL} is readily computed on the lattice, whereas the computation of $A_0$ in QCD is more difficult, and we shall resort to an effective approach here that has been well-tested within functional QCD, \cite{Haas:2013qwp}. We rewrite the effective potential in terms of $L(A_0)$ and $\bar L(A_0)$, 
\begin{align} 
    V_k(\rho,A_0) = V_k(\rho,L,\bar L) =V_{\mathrm{glue},k}(L,\bar L)+V_{\mathrm{mat},k}(\rho,L,\bar L) \,, 
\label{eq:PolpotLbarL}
\end{align} 
using also $V_k$ for the effective potential in terms of $L(A_0)$ in a slight abuse of notation. In its form $V(\rho,L,\bar L)$, the potential is commonly referred to as the Polyakov loop potential. 

It is left to determine the two parts of the effective potential \labelcref{eq:PolpotLbarL}. For the determination of the pure glue potential $V_{\mathrm{glue},k}$, we use the following properties: The glue potentials $V_{\mathrm{glue},k}(A_0)$ in pure Yang-Mills theory and QCD have been computed and related in \cite{Braun:2007bx, Braun:2009gm, Fister:2013bh, Haas:2013qwp}. Importantly, their relation is given by a simple rescaling of the reduced temperature $(T-T_c^\mathrm{glue})/T_c^\mathrm{glue}$ with the phase transition temperature $T_c^\mathrm{glue}$ of the glue potential. The latter should not be confused with the physical crossover temperature $T_c$. This entails that the glue potential in QCD can be easily derived from that in Yang-Mills theory. This simple rescaling follows directly from the flow equation of the glue potential in terms of the first two loops in \Cref{fig:QCD_FlowEquation}: The rescaling is related to that of the ghost and gluon propagators at finite temperature or rather their mass gaps, see also  \cite{Fu:2019hdw}. While the ghost propagator is essentially unchanged, the mass gap of the gluon increases at finite temperature; roughly speaking the mass gap in the vacuum is augmented with a thermal screening mass.

In summary, this inspires a simple representation of the glue potential $V_{\mathrm{glue},k}$, and in the present work we use the same glue potential as in \cite{Fu:2021oaw}, with the parameters of the glue potential being $\alpha=0.53$ and $T_c^\mathrm{glue} =237$\,MeV, for more details see \cite{Fu:2021oaw}. This derivation of the glue potential is benchmarked with the fluctuation observables at $\mu_B=0$ which are in quantitative agreement with the lattice results. This concludes the determination of the glue potential at $\mu_B=0$.

At $\mu_B\neq 0$ we use that the gluon propagator has only a very subleading $\mu_B$-dependence for the baryon chemical potentials under consideration as it is only the quark contributions to the gluon that change. This structural property is confirmed in the explicit computation, see \cite{Fu:2019hdw}. Moreover, the $\mu_B$-dependence of the ghost propagator is suppressed even more, as it only receives $\mu_B$-dependences via the gluon. Consequently, we also consider a $\mu_B$-independent glue potential.

The second term in \labelcref{eq:Vtotal} or \labelcref{eq:PolpotLbarL}, $V_{\textrm{mat},k}$ is the matter part of the potential. Its dominant contribution to the $L\,,\,\bar L$-dependence comes from the quark loop through its direct coupling to $A_0$. In turn, the meson loop has a very sub-dominant dependence on $L\,,\,\bar L$ and mainly contributes to the $\rho$-dependence. The coupling between $q,\bar q$ and the mesonic degrees of freedom $\phi=(\sigma, \bm\pi)$ via $h_k$ in \labelcref{eq:action} leads to a relation between the Polyakov loop  and meson mean fields, carried by the quark loop in the flow equation in \Cref{fig:QCD_FlowEquation}. In the present work we compute the matter part of the potential, $V_{\textrm{mat},k}$, directly via its flow equation given by the last two terms in \Cref{fig:QCD_FlowEquation}, using direct  QCD input.

%%%%%%%%%%%%%%%%%%%%%%%%%%%%%%%%%%%%%%%%%%
\subsubsection{Quark-meson coupling in functional QCD with emergent composites}
\label{app:Yukawa}

In our previous work \cite{Fu:2021oaw}, the two-flavor LEFT was improved towards 2+1 flavors by adapting the scale matching between the 2+1-flavor and two-flavor QCD to the LEFT. This allowed us to study QCD phase transitions and baryon number fluctuations in $2+1$-flavor QCD via the $2$-flavor LEFT. Moreover, most of the low energy dynamics was computed in the LEFT. 

\begin{figure*}[t]
\includegraphics[width=0.45\textwidth]{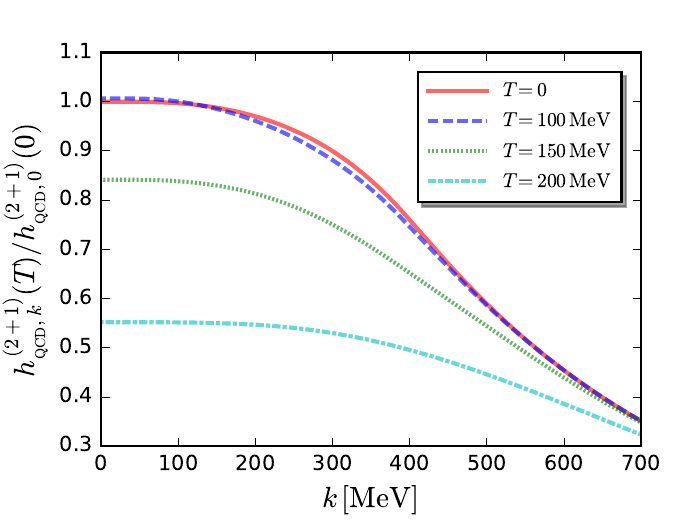}
\caption{Thermal Yukawa coupling, normalised to unity for $k=0$ in the vacuum, as a function of the RG scale $k$ for several temperature values, obtained within the first-principles fRG computation of $2+1$-flavor QCD in \cite{Fu:2019hdw}, see also \labelcref{eq:hQCD}.}\label{fig:hQCD-k}
\end{figure*}

In this work, we adopt a more direct approach that augments the present LEFT with the full QCD dynamics of relevant correlations. Most prominently, we use the thermal dynamics of the Yukawa coupling in  first principles $2+1$-flavor QCD at finite temperature and baryon chemical potential, 
\begin{align}
    h_k(T)=&h_0 \frac{h^{{\scriptscriptstyle \mathrm{QCD}}}_{k}(T)}{h^{\scriptscriptstyle \mathrm{QCD}}_{0}(0)}\,. 
    \label{eq:hQCD}
\end{align}
Here, $h^{{\scriptscriptstyle \mathrm{QCD}}}_{k}(T)$ denotes the scale-dependent bare Yukawa coupling of $2+1$-flavor QCD computed in \cite{Fu:2019hdw}. It is shown for illustration in \Cref{fig:hQCD-k} at several selective values of temperature. The $\mu_B$-dependence of the bare Yukawa coupling is not taken into account, since it is found that this dependence is very small and can be safely neglected \cite{Fu:2019hdw}. The absolute value of the Yukawa coupling is determined by its vacuum value $h_0$ in \labelcref{eq:hQCD}. Together with the initial parameters in the mesonic potential it is determined with observables in the vacuum, see below. 

By solving the flow equations of the effective action in \labelcref{eq:action}, including those of the effective potential and the wave functions, cf. \cite{Fu:2021oaw} for the details, one is able to obtain the thermodynamic potential density, 
\begin{align}
    \Omega[T,\mu_B]=&V_{\textrm{eff}}(\rho,A_0)-c_\sigma\sigma \,, \qquad \textrm{with}\qquad V_{\textrm{eff}}(\rho,A_0) = V_{k=0}(\rho,A_0)\,, 
\end{align}
where $V_{\textrm{eff}}$ is the full  effective potential obtained at the physical scale $k=0$. The flow equations are integrated from an initial UV scale $k=\Lambda$, that is chosen to be the same value $\Lambda=700$\,MeV as in \cite{Fu:2021oaw}. Furthermore, one also has to specify the effective potential at the initial scale, which provides us with the initial conditions for the flow equations. The initial effective potential reads
\begin{align}
    V_{\mathrm{mat},  \Lambda}(\rho)=&\nu_{\Lambda}\rho+\frac{\lambda_{\Lambda}}{2}\rho^2  \,.
    \label{eq:VmatLam}
\end{align}
The parameters $\lambda_{\Lambda}$, $\nu_{\Lambda}$ in \labelcref{eq:VmatLam}, $h_0$ in \labelcref{eq:hQCD} and the strength of explicit chiral symmetry breaking $c_\sigma$ in \labelcref{eq:action} constitute the set of parameters of the QCD-assisted LEFT in this work.

%
%%%%%%%%%%%%%%%%%%%%%%%%%%%%%
%%%%%%%%%%%%%%%%%%%%%%%%%%%%%
%

%
%%%%%%%%%%%%%%%%%%%%%%%%%%%%%
\begin{table*}[t]
  \begin{center}
 \begin{tabular}{cccc}
    \hline\hline & & &   \\[-2ex]   
    & Set I & \hspace{0.5cm} Set II \hspace{0.5cm} &  Set III  \\[1ex]
    \hline & & &  \\[-2ex]
   $\lambda_{\Lambda}$   & 10.15  &10.00 & 10.15  \\[1ex]
   $\nu_{\Lambda}$ $[\mathrm{GeV}^2]$  & 0.53 & 0.59 & 0.39  \\[1ex]
   \hspace{0.5cm}$c_\sigma$ $[\times10^{-3}\mathrm{GeV}^3]$\hspace{0.5cm}   & 1.6 & 1.5 & 1.85 \\[1ex]
   $h_0$  & 11.6 & 12.9 & 9.05 \\[1ex]
    \hline\hline
  \end{tabular}
  \caption{Three different sets of parameters for the QCD-assisted LEFT in this work.} 
  \label{tab:param-3-set}
  \end{center}\vspace{-0.5cm}
\end{table*}
%%%%%%%%%%%%%%%%%%%%%%%%%%%%%
%

%
%%%%%%%%%%%%%%%%%%%%%%%%%%%%%
\begin{table*}[t]
  \begin{center}
 \begin{tabular}{ccccc}
    \hline\hline & & & & \\[-2ex]   
    & Set I & \hspace{0.5cm} Set II \hspace{0.5cm} &  Set III  & \hspace{0.5cm}  QCD ($N_f=2+1$)\\[1ex]
    \hline & & & &\\[-2ex]
  $m_{\pi}$ [MeV]   & 137  &136 & 137  & 137\\[1ex]
  $m_{\sigma}$ [MeV]  & 431 & 443 & 411  & 510\\[1ex]
    $\langle\sigma\rangle$ [MeV]    & 76 & 71 & 91 & 69\\[1ex]
    $m_l$ [MeV]   & 343 & 346 & 346 & 347\\[1ex]
   \hspace{0.5cm}$(T_{_{\mathrm{CEP}}},{\mu_B}_{_{\mathrm{CEP}}})$ [MeV] \hspace{0.5cm}  & (98, 643) & (108, 604) & (94, 704) & (107, 635)\\[1ex]
    \hline\hline
  \end{tabular}
  \caption{Some observables, including the pion mass $m_{\pi}$, sigma mass $m_{\sigma}$, expectation value of the sigma field $\langle\sigma\rangle$, mass of light quarks $m_l$ in the vacuum as well as the location of the critical end point in the phase diagram, obtained in the QCD-assisted LEFT with the three different sets of parameters in \Cref{tab:param-3-set}. The relevant results obtained in $N_f=2+1$-flavor QCD in \cite{Fu:2019hdw} are also shown for comparison.} 
  \label{tab:observ-param-3-set}
  \end{center}\vspace{-0.5cm}
\end{table*}
%%%%%%%%%%%%%%%%%%%%%%%%%%%%%
%

%
%%%%%%%%%%%%%%%%%%%%%%%%%%%%%
\begin{figure*}[t]
\includegraphics[width=0.45\textwidth]{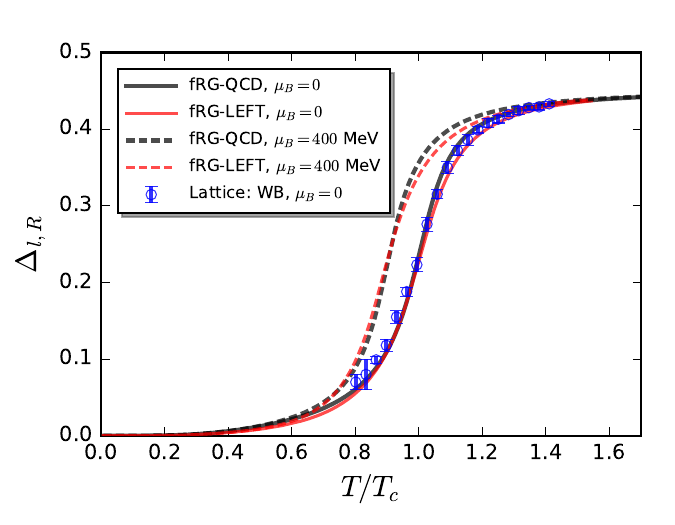}
\caption{Renormalised light quark chiral condensate $\Delta_{l,R}$ as a function of temperature at $\mu_B=0$ and 400\,MeV, calculated in the QCD-assisted fRG-LEFT with the parameters of Set I, in comparison to the results from first-principles $N_f=2+1$-flavor QCD within the fRG approach (fRG-QCD) \cite{Fu:2019hdw} as well as the lattice results at vanishing baryon chemical potential from the Wuppertal-Budapest collaboration (WB) \cite{Borsanyi:2010bp}. Here the pseudo-critical temperature of chiral phase transition at vanishing $\mu_B$ for the $N_f=2+1$-flavor QCD is $T_c=156$\,MeV.}\label{fig:DeltalR}
\end{figure*}
%%%%%%%%%%%%%%%%%%%%%%%%%%%%%
%

%
%%%%%%%%%%%%%%%%%%%%%%%%%%%%%
\begin{figure*}[t]
\includegraphics[width=0.8\textwidth]{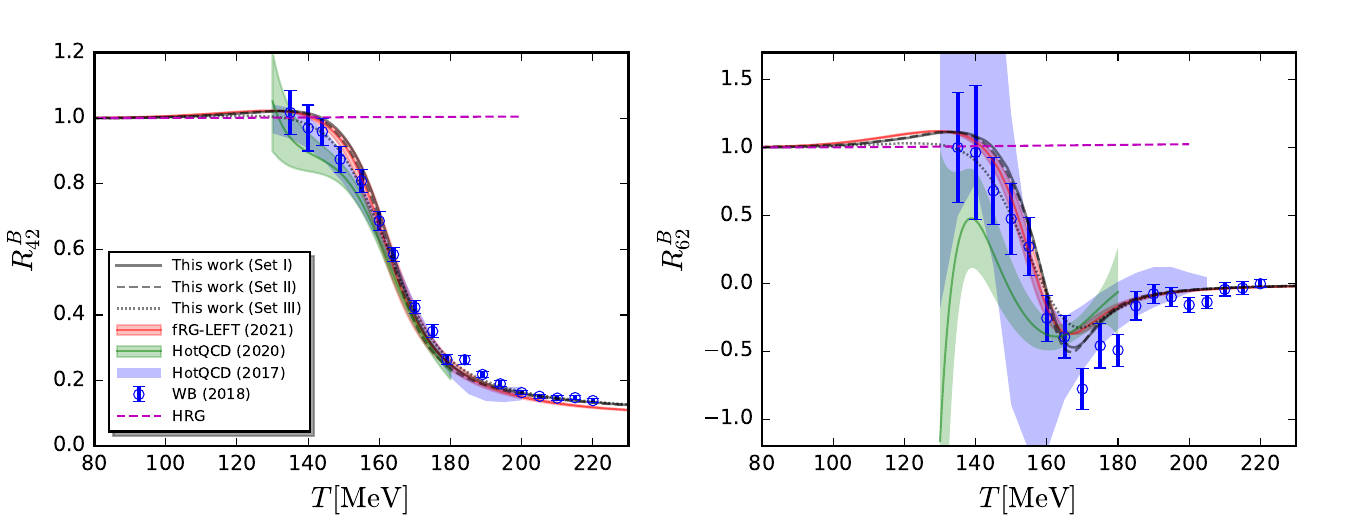}
\caption{$R^{B}_{42}$ (left panel) and $R^{B}_{62}$ (right panel) as functions of the temperature with $\mu_B=0$. Results obtained in this work with the three sets of parameters in \Cref{tab:param-3-set} are compared with our previous results (fRG-LEFT) \cite{Fu:2021oaw}, lattice results from the HotQCD collaboration \cite{Bazavov:2017dus, Bazavov:2017tot, Bazavov:2020bjn} and the Wuppertal-Budapest collaboration (WB) \cite{Borsanyi:2018grb}, and results of hadron resonance gas (HRG) \cite{BraunMunzinger:2003zd}.}\label{fig:R42R62-T-muB0}
\end{figure*}
%%%%%%%%%%%%%%%%%%%%%%%%%%%%%
%

We emphasise that the present QCD-assisted LEFT is significantly improved towards full QCD with the inclusion of QCD dynamics from first-principles calculations within the fRG approach. This is facilitated by the fact, that LEFTs are emerging dynamically within the flow equation approach at low energies and hence can be naturally embedded in full QCD, and systematically improved towards the latter. This embedding and successive improvement also improve the quantitative reliability of LEFT predictions and in consequence also constrain observables such as the location of the CEP.  For instance, the baryon chemical potential for the location of the CEP in the QCD-assisted LEFT is constrained by  
\begin{align}
{\mu_B}_{_{\mathrm{CEP}}}\in (600, 700)\,\textrm{MeV}\,,
\label{eq:CEPLocationQCD-assisted}
\end{align}
which is compatible with the constraints from first-principles functional QCD \cite{Fu:2019hdw, Gao:2020qsj, Gao:2020fbl, Gunkel:2021oya}, see \labelcref{eq:CEPEstimateQCD}. In contradistinction, in the conventional quark-meson (QM) model or Polyakov loop QM model, the CEP is typically located at very high baryon chemical potentials with ${\mu_B}_{_{\mathrm{CEP}}}>$ 850\,MeV but also can be changed very easily. For a collection of  respective results see the recent reviews  \cite{Dupuis:2020fhh, Fu:2022gou}. 

Note also, that this QCD-assisted LEFT also accommodates the considerable uncertainty at high baryon chemical potentials with $\mu_B/T > 4$, where there are still sizable errors for the first-principles QCD calculations with the functional methods \cite{Fu:2019hdw, Gao:2020qsj, Gao:2020fbl, Gunkel:2021oya}. One can utilise this freedom to investigate the dependence of experimental measurements on the theoretical predictions of, e.g., the location of CEP. This information, together with further in-detail studies and increasing wealth of accurate experimental data in this regime, will allow us eventually to pin down the location of CEP as well as the properties of the respective onset regime. 

In \Cref{tab:param-3-set} we show three different sets of parameters of the QCD-assisted LEFT, and the resulting observables are presented in \Cref{tab:observ-param-3-set}, in comparison to the results obtained in the first-principles calculations of $2+1$-flavor QCD \cite{Fu:2019hdw}. Note that parameters of Set I are used in the main text and supplemental material unless stated explicitly.

In \Cref{fig:DeltalR} we compare the renormalised light quark chiral condensate $\Delta_{l,R}$ calculated in the QCD-assisted LEFT in the work with the relevant results obtained from the first-principles $N_f=2+1$-flavor QCD within the fRG approach \cite{Fu:2019hdw} and lattice QCD \cite{Borsanyi:2010bp}. The renormalised light quark chiral condensate is given by
\begin{align}
    \Delta_{l,R}(T,\mu_B) =& -\frac{c_\sigma}{2\mathcal{N}_R}\left[ \sigma_\mathrm{EoM}(T,\mu_B) -\sigma_\mathrm{EoM}(0,0)\right] \,,\qquad\textrm{with} \qquad \left. \frac{\partial V_{\textrm{eff}}(\rho,A_0)}{\partial \sigma}\right|_{\sigma=\sigma_\mathrm{EoM}}=c_\sigma\,. \label{eq:DeltalR}
\end{align}
The field $\sigma_\mathrm{EoM}$ is the constant solution of the equation of motion (EoM), i.e., the expectation value of the sigma field. In \labelcref{eq:DeltalR} the other fields are set to the solutions of their EoMs: $\boldsymbol{\pi}=0$, $q,\bar q=0$ and $A_0=A_0^\textrm{EoM}$. The normalisation constant $\mathcal{N}_R$ is chosen to match the scale in the lattice calculation for the sake of the comparision in \Cref{fig:DeltalR}, see \cite{Fu:2019hdw, Braun:2020ada} for more details. From \Cref{fig:DeltalR} one can see that the results of QCD-assisted LEFT agree well with those obtained from functional QCD and lattice QCD at both vanishing and finite baryon chemical potentials.

In \Cref{fig:R42R62-T-muB0} we show the temperatute dependence of $R^{B}_{42}$ and $R^{B}_{62}$ for vanishing baryon chemical potential, obtained with the three sets of parameters in \Cref{tab:param-3-set}. The results are also compared with our previous results as well as lattice QCD and hadron resonance gas. Importantly, all the results obtained in this work are consistent with lattice results and the results in \cite{Fu:2021oaw}.

%%%%%%%%%%%%%%%%%%%%%%%%%%%%%%%%%
\subsection{Critical end point: Estimates and predictions}
\label{app:CEP-Estimates}

In this Section we explain the QCD estimate for the critical end point or the regime with new phases, 
\begin{align}
600\,\textrm{MeV} \lesssim {\mu_B}_{_{\mathrm{CEP}}}\lesssim 650\, \textrm{MeV}\,,
\label{eq:CEPEstimateQCD-Suppl}
\end{align}
and its systematic error. While \labelcref{eq:CEPEstimateQCD-Suppl} may also be the location of new phases such as an inhomogeneous regime in QCD, in the following we concentrate on the CEP. Moreover, we emphasise that the evaluation of works in the literature is only done by its potential of providing an even qualitative estimate of the critical end point, and the lack of providing such an estimate is by no means a quality judgement of these works, most of which concentrate on other physics phenomena and offer the location of the CEP as added benefit.

It is common in the phenomenological Heavy Ion Collision (HIC) community to provide `scatter' plots of potential CEP positions, democratically including model or LEFT estimates and (functional) QCD computations. Note that we list lattice-inspired estimates with analytic continuation under LEFT estimates, as they do not include the QCD dynamics at finite chemical potential.

This equal-importance evaluation ignores the qualitative difference between LEFTs and functional QCD. The latter includes glue dynamics at finite temperature and chemical potential and LEFTs do not. Functional QCD does this at the cost of approximation of the full diagrammatic system, but it allows for a systematic improvement and the evaluation of apparent convergence of the truncation scheme, for discussions see \cite{Fu:2019hdw} and the reviews \cite{Fischer:2018sdj, Dupuis:2020fhh, Fu:2022gou}. In this spirit we collect the different estimates into groups,
\begin{itemize}
    \item[(A)] Functional QCD,
    \item[(B)] Analytic continuations of lattice QCD results at $\mu_B=0$,
    \item[(C)] Low energy effective theories (LEFTs),
\end{itemize}
where also (B) falls into the LEFT class. We also emphasise that self-consistent functional QCD computations at finite chemical potential $\mu_B\neq 0$ \textit{define} a QCD-based analytic continuation of QCD results at $\mu_B=0$. Hence, if functional QCD predictions agree with the lattice results at $\mu_B=0$, that at $\mu_B\neq 0$ offer the qualitatively best analytic continuation of QCD to date:  
\begin{itemize}
    \item[(1)] They \textit{define} an analytic continuation of QCD at $\mu=0$,
    \item[(2)] This analytic continuation surpasses qualitatively any analytic continuation based on QCD results at $\mu_B=0$ in (B), as they include QCD dynamics at $\mu_B\neq 0$,
    \item[(3)] Functional approaches allow for a systematic improvement of the approximation used within apparent convergence. 
\end{itemize}
In short, functional QCD provides QCD benchmark results at finite density as lattice QCD does so at vanishing density. Having the above in mind, we evaluate CEP estimates in the literature with three criteria \textit{(i)-(iii)}: 
\begin{itemize}
\item[(i)] \textbf{Benchmark results:} Only consider estimates within the approaches (A-C) that meet benchmark results from lattice QCD and functional QCD at $\mu_B=0$ and functional QCD at $\mu_B\neq 0$. In the latter case, only functional QCD results within the quantitatively reliable regime are benchmarks. At present, this is the regime 
\begin{align}
    \frac{\mu_B}{T} \lesssim 4\,. \label{eq:fQCDmuBT}
\end{align}
In the regime \labelcref{eq:fQCDmuBT}, QCD physics is dominated by quarks, gluons and the scalar-pseudoscalar four-quark channels and its multi-scattering processes. Beyond the regime \labelcref{eq:fQCDmuBT}, functional QCD computations indicate the importance of further fluctuations: In \cite{Fu:2019hdw} signatures of a moat regime in QCD, \cite{Pisarski:2021qof}, have been found for $\mu_B/T \gtrsim 4$. In \cite{Braun:2019aow} the dominance of diquark fluctuations for $\mu_B/T \gtrsim 6$ is found. While the latter happens very rapidly, a very conservative estimate also gives  $\mu_B/T \gtrsim 4$ for the regime with a rising importance of diquark fluctuations. Both effects are currently included in systematic improvements of the approximations in functional QCD, both fRG and Dyson-Schwinger equation (DSE) computations, which will push the regime \labelcref{eq:fQCDmuBT} towards larger $\mu_B/T$. 

\item[(ii)] \textbf{State-of-the-art CEPs:} We only consider the newest estimate from (A-C) within a given approach: Old estimates are dropped, if there are newer estimates within the same framework obtained within systematic improvements. 

\item[(iii)] \textbf{CEPs within the range of validity of (B-C):} Parameters in a given LEFT including analytic continuations of lattice results (B-C) are adjusted at some temperature and chemical potential $(T,\mu_B)_\textrm{adj}$, typically in the vacuum, $(T,\mu_B)_\textrm{adj}=(0,0)$ (C) or at finite temperature $(T,\mu_B)_\textrm{adj}=(T,0)$ (B). This entails that estimates for large $(T,\mu_B)-(T,\mu_B)_\textrm{adj}$ have an increasing systematic error as LEFTs (B,C) do not accommodate the QCD dynamics in this regime. Accordingly, estimates from (B,C) for large chemical potentials are not trustworthy.  
\end{itemize}
We now put these criteria to work: As discussed in (i), functional QCD computations indicate the increasing importance of fluctuations beyond gluons, quarks and the scalar-pseudoscalar channel for $\mu_B/T$ beyond \labelcref{eq:fQCDmuBT}. While these effects are partially included in state-of-the-art computations within (A), they are certainly absent in (B,C). Accordingly, we drop all LEFT estimates in (B,C) that show a CEP beyond this regime. 

Moreover, we drop all LEFT and functional QCD estimates  (A,C) that do not meet functional QCD and lattice benchmark results. Specifically, this concerns the curvature of the phase boundary at $\mu_B=0$, 
\begin{align}
    \kappa \approx 0.015(2)\,, \label{eq:kappaBenchmark}
\end{align} 
from the lattice and quantitative functional QCD computations, and the location of the crossover line for $\mu_B/T\lesssim 4$ from quantitative functional QCD computations. With these criteria all LEFT estimates (B,C) have to be dropped, they are well beyond the regime of validity of the respective LEFT including analytic continuations from lattice results at $\mu_B=0$, for a recent  comprehensive survey of $\kappa$'s from functional approaches and lattice QCD, see~\cite{Bernhardt:2023ezo}, Figure~3. Note that the curvature in \labelcref{eq:kappaBenchmark} is the leading-order expansion coefficient of the pseudo-critical temperature for the crossover around $\mu_B=0$, i.e.,
\begin{align}
    \frac{T_c(\mu_B)}{T_c}&=1-\kappa \left(\frac{\mu_B}{T_c}\right)^2+\kappa_4 \left(\frac{\mu_B}{T_c}\right)^4+\cdots\,,\label{eq:curv-defi}
\end{align}
with $T_c=T_c(\mu_B=0)$, where $\kappa_4$ is the next-to-leading-order one, that is found to be significantly smaller than $\kappa$ \cite{Bernhardt:2023ezo}.

This leaves us with the functional QCD results  \cite{Fu:2019hdw, Gao:2020fbl, Gunkel:2021oya}. For example, \cite{Gao:2020qsj} has been left out even though it falls in the regime \labelcref{eq:CEPEstimateQCD-Suppl}, as it is superseded by \cite{Gao:2020fbl}. Moreover, the estimates in \cite{Isserstedt:2019pgx} and those reported in \cite{Fischer:2018sdj} are left out due to a combination of (i) and (ii): The respective $\kappa\approx 0.0238$ does not agree within the systematic error bars in \labelcref{eq:kappaBenchmark}, (i), and the estimate is superceded by that in \cite{Gunkel:2021oya}, (ii). In our opinion, these examples illustrate impressively the ongoing systematic improvement of functional QCD estimates and predictions. 

%
%%%%%%%%%%%%%%%%%%%%%%%%%%%%%
\begin{figure}[t]
\includegraphics[width=0.5\textwidth]{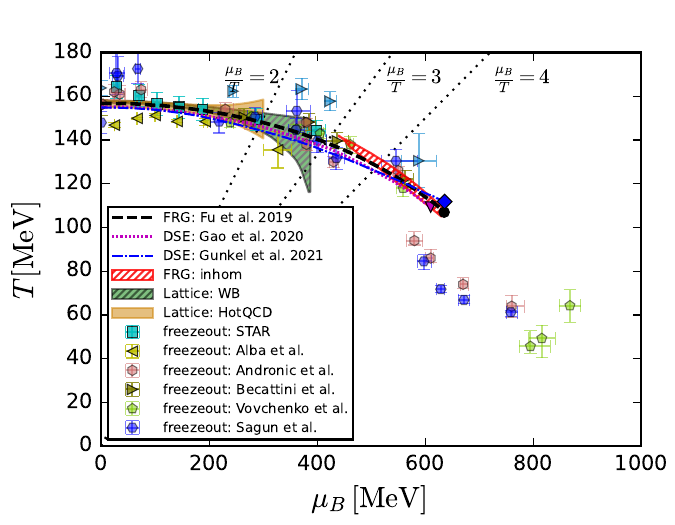}
\caption{Crossover lines from functional QCD and lattice QCD. FRG: Fu \textit{et al.} \cite{Fu:2019hdw}, DSE: Gao \textit{et al.} \cite{Gao:2020fbl}, DSE: Gunkel \textit{et al.} \cite{Gunkel:2021oya}. Lattice: Wuppertal-Budapest Collaboration
\cite{Bellwied:2015rza} (WB) and HotQCD Collaboration \cite{Bazavov:2018mes} (HotQCD). The red hatched area denotes a region of inhomogeneous instability obtained in fRG \cite{Fu:2019hdw}, known also as the moat regime \cite{Pisarski:2021qof}. The estimates of the CEP from functional methods are in the regime \labelcref{eq:CEPEstimateQCD-Suppl}. While all come from functional approaches, the respective approximations and resummations are different, which further improves the respective reliability. We have also added freeze out points: STAR \cite{Adamczyk:2017iwn}, Alba \textit{et al.} \cite{Alba:2014eba}, Andronic \textit{et al.} \cite{Andronic:2017pug}, Becattini \textit{et al.} \cite{Becattini:2016xct}, Vovchenko \textit{et al.} \cite{Vovchenko:2015idt}, Sagun \textit{et al.} \cite{Sagun:2017eye}. Note that freeze-out data from Becattini \textit{et al.} are shown in two different colors, corresponding to with (light blue) and without (dark green) afterburner-corrections, respectively.}
\label{fig:PhaseStructure2023} 
\end{figure}
%%%%%%%%%%%%%%%%%%%%%%%%%%%%%
%

We also comment on the highly interesting development on estimates via analytic continuation of lattice results at $\mu_B=0$: In the past two years these estimates converge towards the regime \labelcref{eq:CEPEstimateQCD-Suppl} singled out by functional QCD \cite{Bollweg:2022rps}. While this has been interpreted as an increasing accuracy of the respective estimates, the situation is more complex. The systematics underlying the analytic continuation results of QCD observables at $\mu_B=0$ implicitly or explicitly relies on the absence or rather the importance of new physics such as a moat regime or diquark fluctuations for baryon chemical potentials smaller than that of the CEP. Moreover, the $\mu_B$-dependence of the crossover line has to be approximated well by the lowest order polynomial in $\mu_B^2$. These assumptions underlie the estimates via an analytic continuation but cannot be proven to be valid for $\mu_B=0$, or $\mu_B^2 < 0$.

On the other hand, functional QCD predicts this behaviour quantitatively for \labelcref{eq:fQCDmuBT}, and estimates it for $\mu_B/T \gtrsim 4$ including the regime \labelcref{eq:CEPEstimateQCD-Suppl}. Indeed, beyond this regime the $(\mu_B^2/T^2)^2$ term in the crossover line gets important due to the size of its prefactor $\kappa_4$. Note that the smallness of $\kappa_4$ is due to the expansion in $\mu_B^2/T^2$, while an expansion 
in $\mu_B^2/(\pi T)^2$ or $\mu_B^2/(2\pi T)^2$ is more natural, as the finite temperature frequency scale is the first Matsubara mode, $2 \pi T$ for bosons and $\pi T$ for fermions. In such an expansion $\kappa_2$ and $\kappa_4$ have the same order of magnitude. 

In summary, we are led to the following combined estimate for the crossover line and the location of the CEP collected in \Cref{fig:PhaseStructure2023}. We emphasise that the figure contains all estimates that pass the combined constraints (i-iii): \cite{Fu:2019hdw, Gao:2020fbl, Gunkel:2021oya}, starting with the first one in \cite{Fu:2019hdw}. The CEP estimates from functional approaches cluster in a very small regime in \labelcref{eq:CEPEstimateQCD-Suppl} which provides further reliability to the estimate. Note that this estimate can be used to inform the parametric Ising-to-QCD mapping equation of state, e.g.~in \cite{Mroczek:2020rpm, Dore:2022qyz}. 

We close with the remark that this combination still only provides an estimate and not a full quantitative prediction: While the computations accommodate part of the dynamics of the potential moat regime, \cite{Pisarski:2021qof}, and diquark fluctuations, a quantitative prediction has to include these effects systematically. We hope to report on such a prediction in the near future.

%%%%%%%%%%%%%%%%%%%%%%%%%%%%%%%%%%
\subsection{Baryon number fluctuations in a grand canonical ensemble}
\label{app:GCE}

%
%%%%%%%%%%%%%%%%%%%%%%%%%%%%%
\begin{figure}[t]
\includegraphics[width=0.8\textwidth]{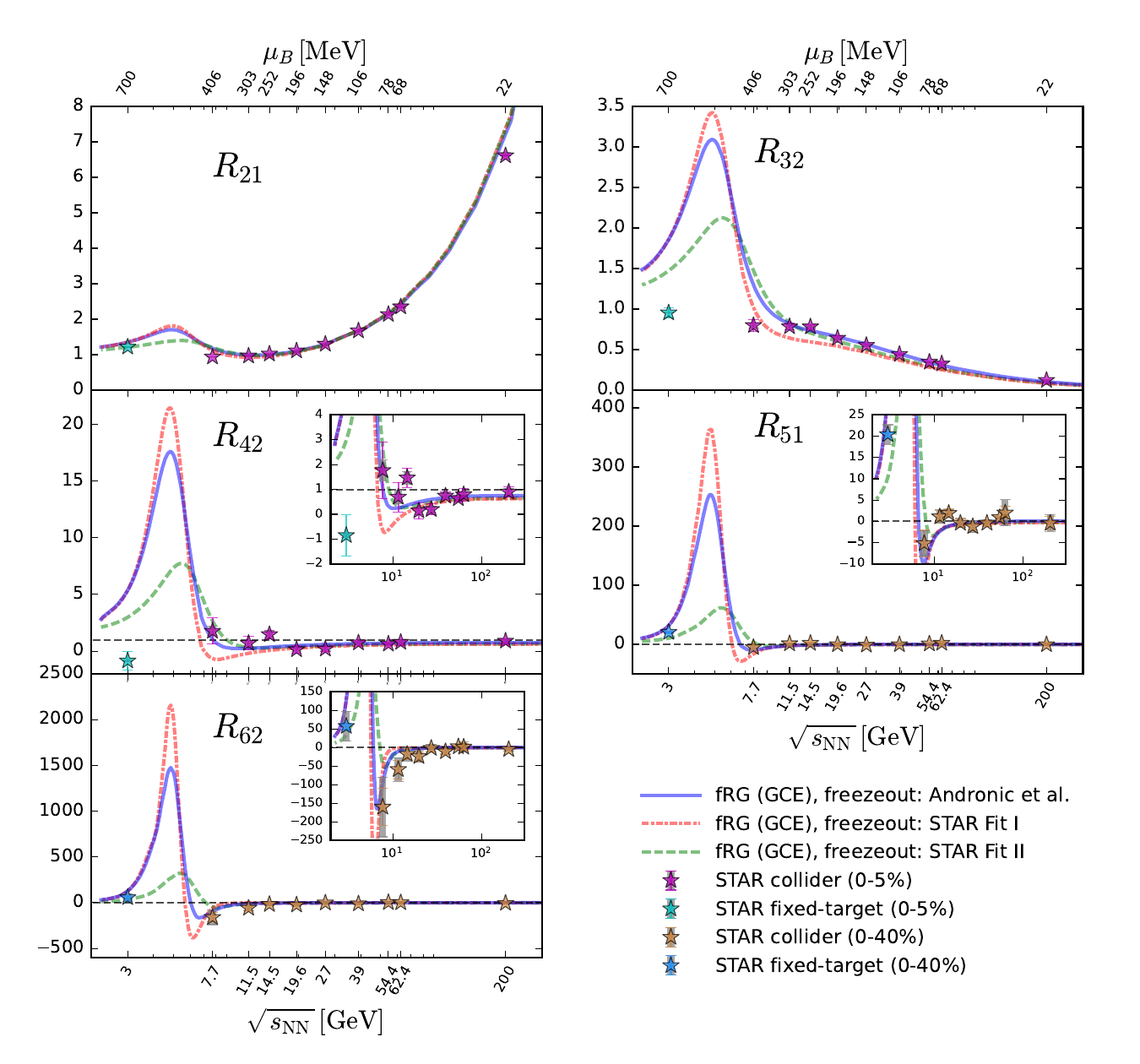}
\caption{Baryon number fluctuations of different orders as functions of the collision energy, calculated in the QCD-assisted LEFT for three different freeze-out curves \cite{Fu:2021oaw} for a grand canonical ensemble, in comparison to STAR results of $R_{21}^p$, skewness $R_{32}^p$, kurtosis $R_{42}^p$ of net-proton distributions for central (0-5\%) collisions \cite{STAR:2020tga}, hyper-order cumulants $R_{51}^p$, $R_{62}^p$ with centrality 0-40\% \cite{STAR:2023esa},   proton cumulants in fixed-target collisions at $\sqrt{s_{\mathrm{NN}}}$=3 GeV \cite{STAR:2021fge}. Inlays show the zoomed-in view.}
\label{fig:chi-GCE}
\end{figure}
%%%%%%%%%%%%%%%%%%%%%%%%%%%%%
%

%
%%%%%%%%%%%%%%%%%%%%%%%%%%%%%
\begin{figure}[t]
\includegraphics[width=0.45\textwidth]{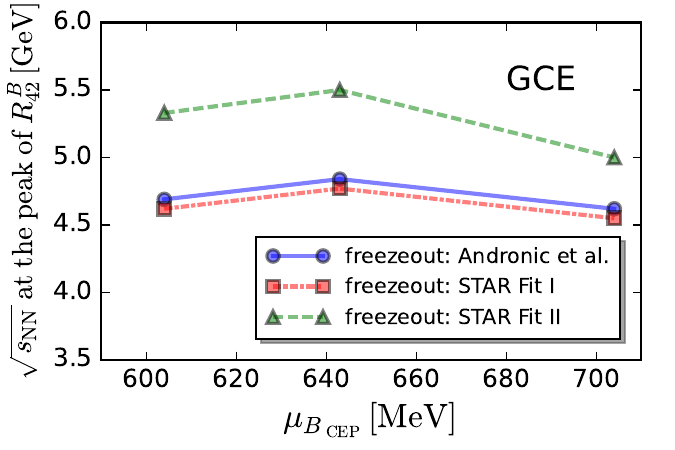}\hspace{0.3cm}
\includegraphics[width=0.45\textwidth]{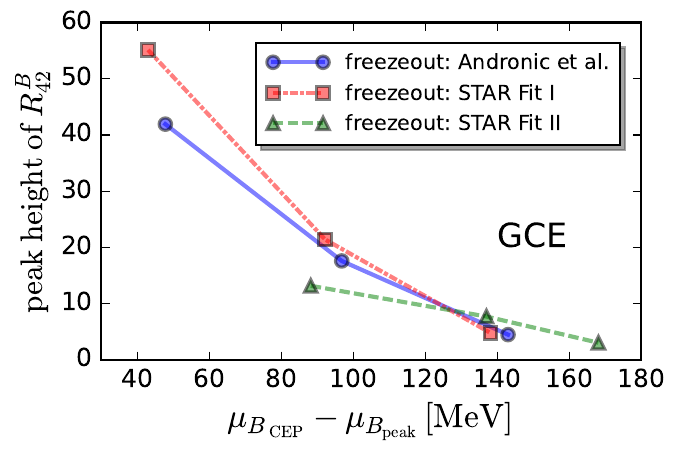}
\caption{Left panel: Dependence of the position of the peak in $R_{42}^{B}$ on the location of the CEP and the freeze-out curve \cite{Fu:2021oaw}, where the computation is done in a grand canonical ensemble. Right panel: Height of the peak in the grand canonical $R_{42}^{B}$ as a function of the difference between the ${\mu_B}_{_{\mathrm{CEP}}}$ and  ${\mu_B}_{_{\mathrm{peak}}}$, where the latter corresponds to $\mu_B$ related to the $\sqrt{s_{\mathrm{NN}}}$ of the peak in $R_{42}^{B}$.}
\label{fig:peak-muBCEP-GCE}
\end{figure}
%%%%%%%%%%%%%%%%%%%%%%%%%%%%%
%

In \Cref{fig:chi-GCE} we show the ratios of baryon number fluctuations $R_{21}^{B}$, $R_{32}^{B}$, $R_{42}^{B}$, $R_{51}^{B}$, $R_{62}^{B}$ in the grand canonical ensemble as functions of the collision energy on the three different freeze-out curves. The CEP is located at $(T_\mathrm{CEP},\mu_{B_{\mathrm{CEP}}})=(98,643)$\,MeV. In comparison to \Cref{fig:chi-CE}, the canonical corrections arising from the global baryon number conservation in the regime of low collision energy are absent in \Cref{fig:chi-GCE}. Evidently, the theoretical results of fluctuations in a grand canonical ensemble are well consistent with experimental data in the regime of collision energy $\sqrt{s_{\mathrm{NN}}}\gtrsim$ 11.5 GeV, but the agreement is worsening with the decreasing collision energy, in particular for $\sqrt{s_{\mathrm{NN}}}\lesssim$ 10 GeV. Moreover, in the regime of low collision energy the magnitudes of high-order fluctuations, in particular the hyper-order ones, in the grand canonical ensemble are significantly larger than those with canonical corrections. The peak structure in the energy range between 3 and 7.7 GeV is still there in the grand canonical ensemble, even more prominent. In \Cref{fig:peak-muBCEP-GCE} we show the grand canonical results for the dependence of the peak position and peak height on the location of the CEP and the freeze-out curve. Similar with the canonical results in \Cref{fig:peak-muBCEP}, we find that the height of peak depends on the location of the CEP, and its position is more sensitive to the freeze-out curve. The baryon chemical potential related to the peak, i.e., ${\mu_B}_{_{\mathrm{peak}}}$, is relatively larger in the grand canonical ensemble than that in the canonical one.

%%%%%%%%%%%%%%%%%%%%%%%%%%%%%%%%%%%%%%%%%%%%%%%%
\subsection{Canonical corrections from global baryon number conservation}
\label{app:canonical}

We use the subensemble acceptance method (SAM) proposed in \cite{Vovchenko:2020tsr} to encode the canonical corrections to the grand canonical fluctuations, arising from the effects of global baryon number conservation in the regime of low collision energy. In SAM, the subensemble volume measured in the acceptance window reads $V_1=\alpha\, V$, where $V$ is the volume of the whole system, and $\alpha$ stands for the ratio between them. It is found in \cite{Vovchenko:2020tsr} that the cumulants and their ratios measured in the sub-system can be related to the grand canonical fluctuations through 
\begin{align}
    \bar R_{21}^B=&\beta R_{21}^B\,,\qquad \bar R_{32}^B= (1-2\alpha)R_{32}^B\,,\qquad
    \bar R_{42}^B=(1-3\alpha \beta)R_{42}^B-3\alpha \beta (R_{32}^B)^2\,,\\[2ex]
    \bar R_{51}^B=&\beta (1-2\alpha)(1-2\alpha \beta)R_{51}^B-10 \alpha \beta^2(1-2\alpha)R_{42}^B R_{32}^B R_{21}^B\,,\label{}\\[2ex]
    \bar R_{62}^B=&\left[1-5\alpha \beta(1-\alpha \beta)\right]R_{62}^B-15\alpha \beta(1-3\alpha \beta)R_{51}^B R_{32}^B(R_{21}^B)^{-1}-10\alpha \beta(1-2\alpha)^2(R_{42}^B)^2+45(\alpha \beta)^2 R_{42}^B (R_{32}^B)^2\nonumber\\[2ex]
                   &-15(\alpha \beta)^2(R_{32}^B)^4\,,
\end{align}
with $\beta=1-\alpha$. Here we have used $\bar R_{nm}^B$ to denote the fluctuations including canonical corrections, to be distinguished from the grand canonical $R_{nm}^B$. Obviously, from the equations above one arrives at $\bar R_{nm}^B=R_{nm}^B$ for $\alpha \to 0$: the canonical correction is negligible when the acceptance window is quite smaller than the total system, which is the case in the regime of large collision energy.

%%%%%%%%%%%%%%%%%%%%%%%%%%%%%%%%%%%%%%%%%%%%%%%%
\subsection{Imprint of the location of the CEP on the baryon number fluctuations}
\label{app:locCEP}

%
%%%%%%%%%%%%%%%%%%%%%%%%%%%%%
\begin{figure*}[t]
\includegraphics[width=0.45\textwidth]{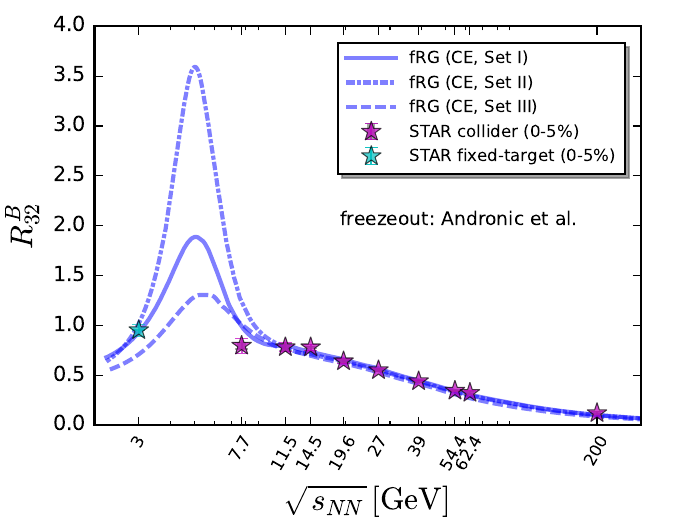}\hspace{0.3cm}
\includegraphics[width=0.45\textwidth]{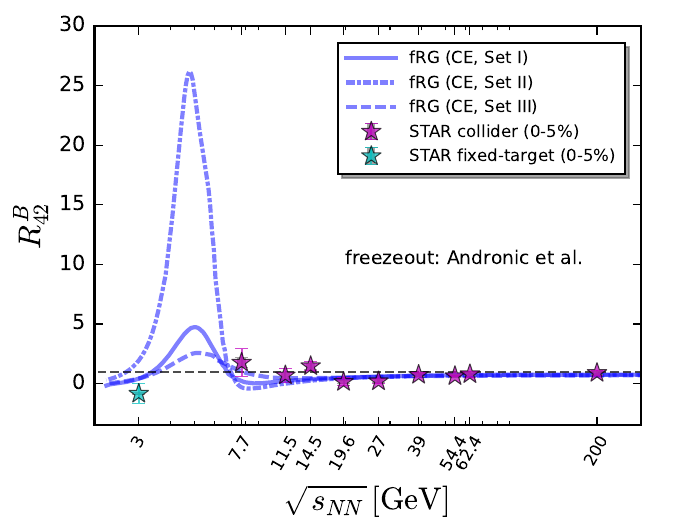}
\caption{Skewness $R_{32}^{B}$ (left panel)  and kurtosis $R_{42}^{B}$ (right panel) of baryon number fluctuations as functions of the collision energy, calculated in the QCD-assisted fRG-LEFT with the three sets of parameters in \Cref{tab:param-3-set} on the freeze-out curve from Andronic {\it et al.} \cite{Andronic:2017pug}. The effects of global baryon number conservation are taken into account in the range of $\sqrt{s_{\mathrm{NN}}}\lesssim$\,11.5\,GeV through the subensemble acceptance method \cite{Vovchenko:2020tsr}. The results are compared with STAR results of $R_{32}^p$ and $R_{42}^p$ of net-proton distributions for central (0-5\%) collisions \cite{STAR:2020tga} and those in fixed-target collisions at $\sqrt{s_{\mathrm{NN}}}$=3 GeV \cite{STAR:2021fge}.}\label{fig:R32R42-sqrtS-CEP-CE}
\end{figure*}
%%%%%%%%%%%%%%%%%%%%%%%%%%%%%
%

%
%%%%%%%%%%%%%%%%%%%%%%%%%%%%%
\begin{figure*}[t]
\includegraphics[width=0.45\textwidth]{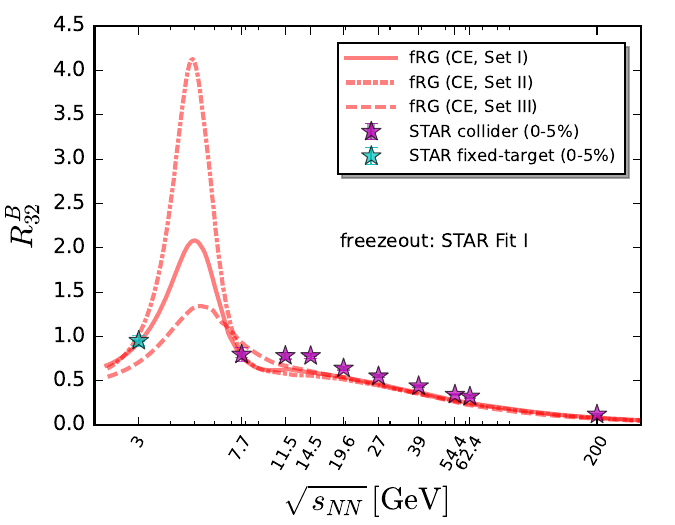}\hspace{0.3cm}
\includegraphics[width=0.45\textwidth]{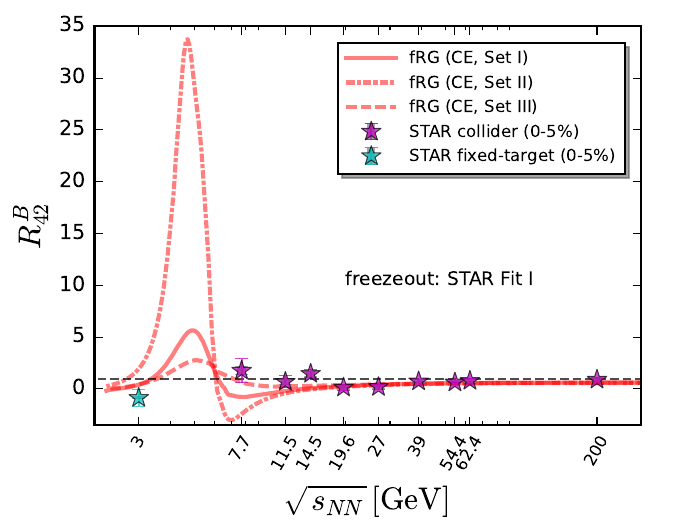}
\caption{Skewness $R_{32}^{B}$ (left panel)  and kurtosis $R_{42}^{B}$ (right panel) of baryon number fluctuations as functions of the collision energy, calculated in the QCD-assisted fRG-LEFT with the three sets of parameters in \Cref{tab:param-3-set} on the freeze-out curve of STAR Fit I \cite{Fu:2021oaw}. The effects of global baryon number conservation are taken into account in the range of $\sqrt{s_{\mathrm{NN}}}\lesssim$\,11.5\,GeV through the subensemble acceptance method \cite{Vovchenko:2020tsr}. The results are compared with STAR results of $R_{32}^p$ and $R_{42}^p$ of net-proton distributions for central (0-5\%) collisions \cite{STAR:2020tga} and those in fixed-target collisions at $\sqrt{s_{\mathrm{NN}}}$=3 GeV \cite{STAR:2021fge}.}\label{fig:R32R42-sqrtS-CEP-CE-CFStI}
\end{figure*}
%%%%%%%%%%%%%%%%%%%%%%%%%%%%%
%

%
%%%%%%%%%%%%%%%%%%%%%%%%%%%%%
\begin{figure*}[t]
\includegraphics[width=0.45\textwidth]{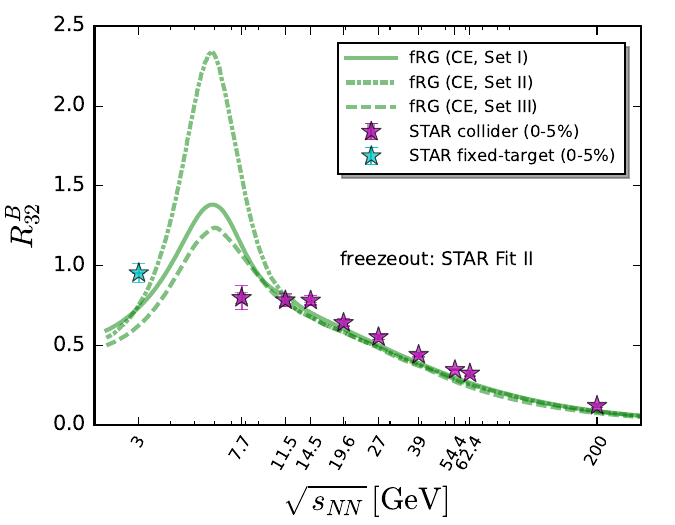}\hspace{0.3cm}
\includegraphics[width=0.45\textwidth]{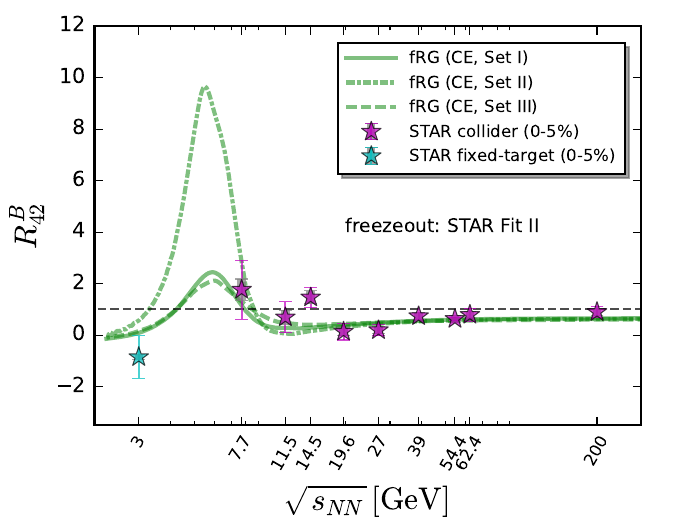}
\caption{Skewness $R_{32}^{B}$ (left panel)  and kurtosis $R_{42}^{B}$ (right panel) of baryon number fluctuations as functions of the collision energy, calculated in the QCD-assisted fRG-LEFT with the three sets of parameters in \Cref{tab:param-3-set} on the freeze-out curve of STAR Fit II \cite{Fu:2021oaw}. The effects of global baryon number conservation are taken into account in the range of $\sqrt{s_{\mathrm{NN}}}\lesssim$\,11.5\,GeV through the subensemble acceptance method \cite{Vovchenko:2020tsr}. The results are compared with STAR results of $R_{32}^p$ and $R_{42}^p$ of net-proton distributions for central (0-5\%) collisions \cite{STAR:2020tga} and those in fixed-target collisions at $\sqrt{s_{\mathrm{NN}}}$=3 GeV \cite{STAR:2021fge}.}\label{fig:R32R42-sqrtS-CEP-CE-CFStII}
\end{figure*}
%%%%%%%%%%%%%%%%%%%%%%%%%%%%%
%

%
%%%%%%%%%%%%%%%%%%%%%%%%%%%%%
\begin{figure*}[t]
\includegraphics[width=0.45\textwidth]{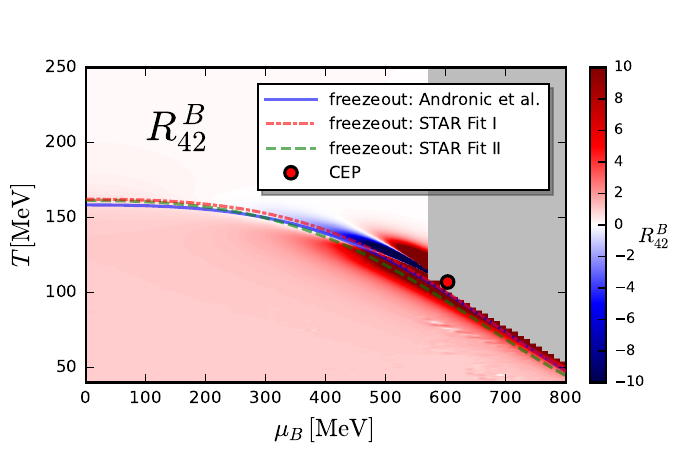}\hspace{0.3cm}
\includegraphics[width=0.45\textwidth]{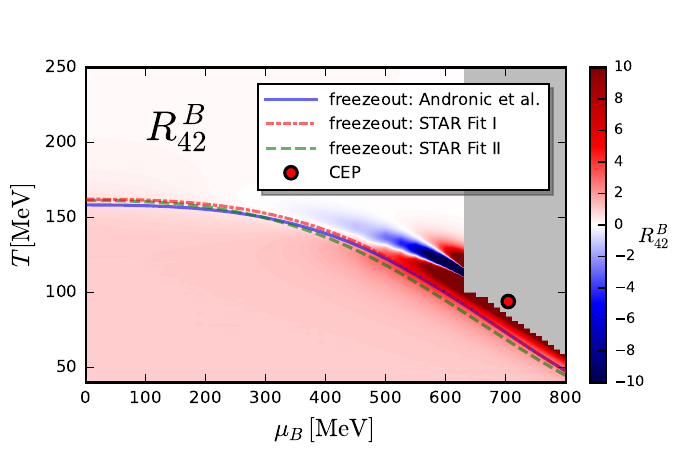}
\caption{Heatmap of $R_{42}^{B}$ in the phase diagram obtained in the QCD-assisted fRG-LEFT for a grand canonical ensemble with the parameters of Set II (left panel) , corresponding to the location of CEP with $(T_{_{\mathrm{CEP}}},{\mu_B}_{_{\mathrm{CEP}}})=(108, 604)$\,MeV, and those of Set III (right panel) with $(T_{_{\mathrm{CEP}}},{\mu_B}_{_{\mathrm{CEP}}})=(94, 704)$\,MeV. Three different freeze-out curves are depicted \cite{Fu:2021oaw}. The gray area shows the region where computation of $R_{42}^{B}$ or $R_{62}^{B}$ is inaccessible within the present numerical set-up.
}\label{fig:R42-phasediagram-CEP600-700}
\end{figure*}
%%%%%%%%%%%%%%%%%%%%%%%%%%%%%
%

In this section we discuss the imprint, that the location of CEP in the QCD phase diagram leaves on the collision-energy dependence of  baryon number fluctuations. For this analysis we evaluate the results for baryon number fluctuations for  three different locations of the CEP in the QCD-assisted LEFT in \Cref{tab:observ-param-3-set} within the regime \labelcref{eq:CEPLocationQCD-assisted}. 

In \Cref{fig:R32R42-sqrtS-CEP-CE} we show $R_{32}^{B}$ and $R_{42}^{B}$ as functions of the collision energy obtained in the QCD-assisted fRG-LEFT with the three different sets of parameters in \Cref{tab:param-3-set}, related to the different locations of CEP as shown in \Cref{tab:observ-param-3-set}. The computation is done on the freeze-out curve from Andronic {\it et al.} \cite{Andronic:2017pug}. Relevant results calculated on the freeze-out curves of STAR Fit I and STAR Fit II are presented in \Cref{fig:R32R42-sqrtS-CEP-CE-CFStI} and \Cref{fig:R32R42-sqrtS-CEP-CE-CFStII}, respectively. The effects of global baryon number conservation are taken into account in the range of $\sqrt{s_{\mathrm{NN}}}\lesssim$\,11.5\,GeV through the SAM by employing the same approach as that in  \Cref{fig:chi-CE}. One can see that the height of the peak both in $R_{32}^{B}$ and $R_{42}^{B}$ in the energy range of fixed-target mode with 3 GeV $\lesssim\sqrt{s_{\mathrm{NN}}}\lesssim$ 7.7 GeV decreases with the increasing ${\mu_B}_{_{\mathrm{CEP}}}$. This can be readily understood from the heatmap of $R_{42}^{B}$ in the phase diagram in \Cref{fig:R42-phasediagram-CEP600-700} as well as in \Cref{fig:R42R62-phasediagram}. In \Cref{fig:R42-phasediagram-CEP600-700} $R_{42}^{B}$ is calculated in the phase diagram with a CEP located at $(T_{_{\mathrm{CEP}}},{\mu_B}_{_{\mathrm{CEP}}})=(108, 604)$\,MeV and $(T_{_{\mathrm{CEP}}},{\mu_B}_{_{\mathrm{CEP}}})=(94, 704)$\,MeV, respectively. Combining them with the top panel of \Cref{fig:R42R62-phasediagram}, where one has $(T_\mathrm{CEP},\mu_{B_{\mathrm{CEP}}})=(98,643)$\,MeV, one observes that as the CEP moves rightward, it deviates more and more away from the freeze-out curves, which results in the decrease of fluctuations. In turn, this property provides us with the possibility to infer the location of CEP by measuring the height of peak in experiments in the future. In contrast, the position of peak is only mildly influenced by the location of CEP in the region of ${\mu_B}_{_{\mathrm{CEP}}}$ investigated, but rather relatively sensitive on the freeze-out curves.

%\end{widetext}

\end{document}